\documentclass{aa}  

\usepackage{graphicx}
\usepackage{txfonts}
\usepackage[colorlinks=true,allcolors=blue]{hyperref}

\usepackage{pdflscape}
\usepackage{subcaption}
\usepackage{threeparttable}

\usepackage{xspace}
\newcommand{\kms}{km~s$^{-1}$\xspace}
\newcommand{\ergs}{erg~s$^{-1}$\xspace}
\newcommand{\Msunyr}{M$_{\odot}$ yr$^{-1}$\xspace}

\newcommand{\hb}{H$\beta$\xspace}
\newcommand{\ha}{H$\alpha$\xspace}

\newcommand{\oii}{[O\,{\sc{ii}}]\xspace}

\newcommand{\sivi}{[Si\,{\sc{vi}}]\xspace}
\newcommand{\siten}{[Si\,{\sc{x}}]\xspace}

\newcommand{\siii}{[S\,{\sc{iii}}]\xspace}
\newcommand{\sii}{[S\,{\sc{ii}}]\xspace}
\newcommand{\pii}{[P\,{\sc{ii}}]\xspace}
\newcommand{\nii}{[N\,{\sc{ii}}]\xspace}

\newcommand{\feii}{[Fe\,{\sc{ii}}]\xspace}

\newcommand{\hei}{He\,{\sc{i}}\xspace}
\newcommand{\mgiv}{[Mg\,{\sc{iv}}]\xspace}
\newcommand{\hi}{H\,{\sc{i}}\xspace}
\newcommand{\arvi}{[Ar\,{\sc{vi}}]\xspace}

\newcommand{\ci}{[C\,{\sc{i}}]\xspace}

\newcommand{\Paa}{Pa$\alpha$\xspace}
\newcommand{\Pab}{Pa$\beta$\xspace}
\newcommand{\Pag}{Pa$\gamma$\xspace}
\newcommand{\Pad}{Pa$\delta$\xspace}
\newcommand{\Pae}{Pa$\epsilon$\xspace}

\newcommand{\Brb}{Br$\beta$\xspace}
\newcommand{\Brg}{Br$\gamma$\xspace}
\newcommand{\Brd}{Br$\delta$\xspace}
\newcommand{\Bre}{Br$\epsilon$\xspace}
\newcommand{\Brten}{Br\,{\sc{10}}\xspace}

\newcommand{\Pfg}{Pf$\gamma$\xspace}
\newcommand{\Pfd}{Pf$\delta$\xspace}
\newcommand{\Pfe}{Pf$\epsilon$\xspace}

\newcommand{\Hud}{Hu$\delta$\xspace}

\newcommand{\Ne }{n$_e$}

\begin{document}

   \title{No evidence of AGN features in the nuclei of Arp 220 from JWST/NIRSpec IFS}



   \author{Michele Perna
          \inst{\ref{iCAB}}\thanks{e-mail: mperna@cab.inta-csic.es}
          \and
          Santiago Arribas\inst{\ref{iCAB}}
          \and
          Isabella Lamperti\inst{\ref{iCAB}}
          \and 
          Miguel Pereira-Santaella\inst{\ref{iIFF}}
          \and 
          Lorenzo Ulivi\inst{\ref{iUNIFI}, \ref{iOAA}, \ref{iUNITR}}
          \and
          Torsten B{\"o}ker\inst{\ref{iESOba}}
          \and
          Roberto Maiolino\inst{\ref{iKav},\ref{iCav}, \ref{iUCL}}
          \and
          Andrew J. Bunker\inst{\ref{iOxf}}
          \and
          St\'ephane Charlot\inst{\ref{iSor}}
          \and
          Giovanni Cresci\inst{\ref{iOAA}}
          \and
          Bruno Rodr\'iguez Del Pino\inst{\ref{iCAB}}
          \and
          Francesco D'Eugenio\inst{\ref{iKav},\ref{iCav}}
          \and
          Hannah \"Ubler\inst{\ref{iKav},\ref{iCav}}
          \and
          Katja Fahrion\inst{\ref{iESAne}}
          \and
          Matteo Ceci\inst{\ref{iUNIFI}, \ref{iOAA}}
          }

   \authorrunning{M. Perna et al.}
   \institute{
    Centro de Astrobiolog\'ia (CAB), CSIC--INTA, Cra. de Ajalvir Km.~4, 28850 -- Torrej\'on de Ardoz, Madrid, Spain\label{iCAB}
    \and
    Instituto de F\'isica Fundamental, CSIC, Calle Serrano 123, 28006 Madrid, Spain\label{iIFF}
    \and 
    Università di Firenze, Dipartimento di Fisica e Astronomia, via G. Sansone 1, 50019 Sesto F.no, Firenze, Italy\label{iUNIFI}
    \and 
    INAF - Osservatorio Astrofisico di Arcetri, Largo E. Fermi 5, I-50125 Firenze, Italy\label{iOAA}
    \and 
    University of Trento, Via Sommarive 14, I-38123 Trento, Italy\label{iUNITR}
    \and
    European Space Agency, c/o STScI, 3700 San Martin Drive, Baltimore, MD 21218, USA\label{iESOba}
    \and
    Kavli Institute for Cosmology, University of Cambridge, Madingley Road, Cambridge, CB3 0HA, UK\label{iKav}
    \and
    Cavendish Laboratory - Astrophysics Group, University of Cambridge, 19 JJ Thomson Avenue, Cambridge, CB3 0HE, UK\label{iCav}
    \and
    Department of Physics and Astronomy, University College London, Gower Street, London WC1E 6BT, UK\label{iUCL}    
    \and
    Department of Physics, University of Oxford, Denys Wilkinson Building, Keble Road, Oxford OX1 3RH, UK\label{iOxf}
    \and
    Sorbonne Universit\'e, CNRS, UMR 7095, Institut d’Astrophysique de Paris, 98 bis bd Arago, 75014 Paris, France\label{iSor} 
    \and
    European Space Agency, European Space Research and Technology Centre, Keplerlaan 1, 2201 AZ Noordwijk, the Netherlands\label{iESAne}
    }


 
  \abstract
   {Arp 220 is the nearest ultra-luminous infrared galaxy;  it shows evidence of 100 pc-scale molecular outflows likely connected with galaxy-scale outflows traced by ionised and neutral gas. The two highly obscured nuclei of Arp 220 are the site of intense star formation, with extreme (far-infrared based) star-formation rate surface densities, $\Sigma_{SFR} \gtrsim 10^3$ \Msunyr kpc$^{-2}$. Despite extensive investigations searching for active galactic nucleus (AGN) activity in the Arp 220 nuclei, direct evidence remains elusive.}
   {We present JWST/NIRSpec IFS observations covering the $0.9-5.1\ \mu$m wavelength range of the innermost ($5\arcsec \times 4\arcsec$, i.e. $1.8\times 1.5$ kpc) regions of Arp 220. The primary goal is to investigate the potential presence of AGN signatures in the nuclear regions by analysing the spectra extracted from circular apertures of radius 55 pc (0.15\arcsec) around each of the two nuclei.
   }
   {The analysis aims to identify highly ionised gas emission lines (with ionisation potential $> 54$ eV) and other spectral features indicative of AGN activity. Ionised and molecular gas kinematics are also taken into account to study the outflow signatures at $< 60$ pc scales.}
   {We identify $\sim 70$ ionised and $\sim 50$ molecular emission lines in the nuclear spectra of Arp 220. We use recombination line ratios to measure 
   optical extinctions in the range A$_V \sim 11-14$ mag. High ionisation lines are not detected, except the \mgiv line at 4.49 $\mu$m which we interpret as due to shocks rather than to AGN ionisation. We identify broadening and multiple kinematic components in the \hi and H$_2$ lines caused by outflows and shocks, with velocities up to $\sim 550$ \kms. Significantly higher velocities (up to $\sim 900$ \kms) are detected in the off-nuclear regions; however, they do not conclusively represent evidence for AGN activity. 
    }
   {Even with the unprecedented sensitivity of JWST/NIRSpec IFS, achieving an unambiguous identification or exclusion of the presence of an AGN in the Arp 220 system remains challenging, because of its extreme dust obscuration.}

   \keywords{galaxies: individual: Arp 220 -- galaxies: active  -- galaxies: starburst -- galaxies: ISM }

   \maketitle
%

\section{Introduction}

Arp 220 is the closest ultra-luminous infrared galaxy (ULIRG), with a distance of $\sim 78$ Mpc and log $L_{IR}/L_\odot$ = 12.2 (e.g. \citealt{Pereira2021}). This late-stage merger shows several tidal structures in the outskirts (e.g. \citealt{Arp1966, Arribas2001, Perna2020}) and contains two compact ($< 150$ pc), highly obscured nuclei separated by $\sim 370$ pc (1\arcsec) that dominate the infrared luminosity (e.g. \citealt{Scoville2007}).

The nuclear region of Arp 220 is the site of an intense star formation, with a star formation rate SFR $\sim 200 - 250$ \Msunyr (e.g. \citealt{Nardini2010,Varenius2016}) and incredibly high star formation rate surface densities ($\Sigma_{SFR} \sim 10^3-10^4$ \Msunyr kpc$^2$, from radio and far-infrared observations; e.g. \citealt{BarcosMunoz2015,Pereira2021}). Long-term monitoring with very long baseline interferometry has revealed dozens of supernovae and supernova remnants resulting from the extreme star formation occurring in the compact nuclei (e.g. \citealt{Varenius2019}). The off-nuclear regions appear to have post-starburst properties, with weak star formation produced in a few young and low-mass clusters (e.g. \citealt{Perna2020,Chandar2023}), as
expected for a late-stage merger.

Although Arp 220 has been extensively studied from radio to hard X-ray and $\gamma$-ray wavelengths, there is still no convincing direct evidence for the presence of active galactic nucleus (AGN) activity in either of the two nuclei; 
however, indirect arguments provide suggestive indications (e.g. \citealt{Teng2015, Paggi2017, Yoast2019}). 
Accreting super-massive black holes (SMBHs) could be buried in Compton-thick absorbers (i.e. $N_H > 1.5\times 10^{24}$ cm$^{-2}$) at the position of the two nuclei (\citealt{Teng2015, Sakamoto2021a}) and, for instance, explain the excess of $\gamma$-ray flux in comparison to all other diagnostics of star-forming activity (\citealt{YoastHull2017}).

Evidence of multi-phase (i.e. molecular, neutral, and ionised) outflows in Arp 220 have been inferred from the analysis of multi-wavelength observations. Collimated outflows have been discovered in both nuclei on scales of $\sim 100$ pc using, for instance, ALMA observations of carbon monoxide transitions (e.g. \citealt{Wheeler2020,Ueda2022,Lamperti2022}); large-scale outflows have also been identified with optical integral field instruments (IFS; e.g. \citealt{Arribas2001, Colina2004,Perna2020}) and associated with extended soft X-ray emission (e.g. \citealt{McDowell2003}). The possible connection between these multi-phase and multi-scale outflows has not yet been investigated in detail. Consequently, the mechanisms at the origin of these outflows, and hence whether they are due to starburst or AGN winds, are still debated.

In this work, we revisit Arp 220 and its nuclear regions taking advantage of JWST observations performed with the IFS unit of the NIRSpec instrument (\citealt{Jakobsen2022, Boker2022}); these data cover the near-infrared (NIR) wavelength range between 0.95 and 5.1 $\mu$m. The observations were taken as part of the JWST/NIRSpec IFS Guaranteed Time Observations (GTO) survey ``Resolved structure and kinematics of the nuclear regions of nearby galaxies'' (program lead: Torsten B{\"o}ker). The analysis presented in this paper aims to identify AGN signatures in the Arp 220 nuclei, as the presence of highly ionised gas and extreme kinematics unambiguously associated with accreting SMBHs. 

The rest of this paper is organised as follows. Section \ref{sec:reduction} presents the NIRSpec IFS observations
and data reduction. Section \ref{sec:analysis} describes the spectral fitting analysis and the identification of emission line species.  Section \ref{sec:results} reports a qualitative description of the nuclear spectra as well as that of a nearby bright stellar cluster, and some general properties of the interstellar medium (ISM). The multi-phase outflow properties, the absence of highly ionised species, and the comparison between \hi and far-IR SFR tracers are discussed in  Section \ref{sec:discussion}. Finally, Section \ref{sec:conclusions} summarises our conclusions. 
Throughout this work, we assume $\Omega_m=0.286$ and $H_0=69.9$ km/s/Mpc \citep{Bennett2014}. With this cosmology, $1''$ corresponds to a distance of 0.368 kpc at $z=0.018$.

\section{Observations and data reduction}\label{sec:reduction}

Arp 220 was observed in NIRSpec IFS mode as part of the programme 1267 (PI: D. Dicken). The present observations were combined in a single proposal with independent MIRI acquisitions of the same target, with the aim of saving the telescope slew overhead. The observations were collected on 6 March 2023, using the NIRSpec IFS mode (\citealt{Boker2022}). 

The dataset consists of two distinct sets of four-point medium cycling dithering pattern with 15 groups per exposure. The first set is centred at the mid-point between the coordinates of the western (W) and the eastern (E) nuclei; the second set is offset $\sim 1\arcsec$ towards the north-west to cover the shell of \ha and \nii emission (e.g. \citealt{Lockhart2015, Perna2020}). 
The total integration time is 933 seconds per set, in each of the three high resolution (R $\sim 2700$) grating settings, to cover the spectral range from 0.95 to 5.1 $\mu$m. 

\subsection{Data reduction}

We used the STScI pipeline v1.8.2 with CRDS context 1063. A patch was included to correct some important bugs that affect this specific version of the pipeline \citep[see details in][]{Perna2023a}. In particular, we corrected the count-rate images for $1/f$ noise by fitting a polynomial. During stage 2, we masked pixels at the edge of the slices (one pixel wide) to conservatively exclude pixels for which the correction for the throughput of the spectrograph optics (contained in the SFLAT reference files) is unreliable. We also implemented the outlier rejection of \cite{DEugenio2023}. The combination of the eight dithers (with drizzle weighting) allowed us to sub-sample the detector pixels, resulting in cube spaxels with a size of 0.05\arcsec (corresponding to $\sim 20$ pc per spaxel). 

The resulting data cubes are known to show sinusoidal modulations in single spaxel spectra, caused by the undersampling of the point-spread function (PSF). To correct these so-called ``wiggles'', we applied the methodology presented in \cite{Perna2023a}, and adapted to the Arp 220 dataset as reported in \citealt{Uliviinprep}. However, we note that these wiggles do not affect the results presented in this work, because we use spatially integrated spectra, which are much less affected by PSF undersampling (e.g. \citealt{Law2023}). 

In addition to the Arp 220 dataset, in this manuscript we present the JWST/NIRSpec observations of another extra-galactic source, the nearby LIRG 
VV114 (programme 1328, PI: L. Armus; e.g. \citealt{Rich2023}). They were reduced following the same procedure reported above.  

Therefore, for both Arp 220 and VV114 we obtained three data cubes, corresponding to the grating settings G140H/F100LP (referred to as band1 hereinafter), G235H/F170LP (band2), and G395H/F290LP (band3), covering the wavelength range 0.95 -- 5.1 $\mu$m.

\begin{figure}[!tb]
\centering 
\includegraphics[width=0.45\textwidth]{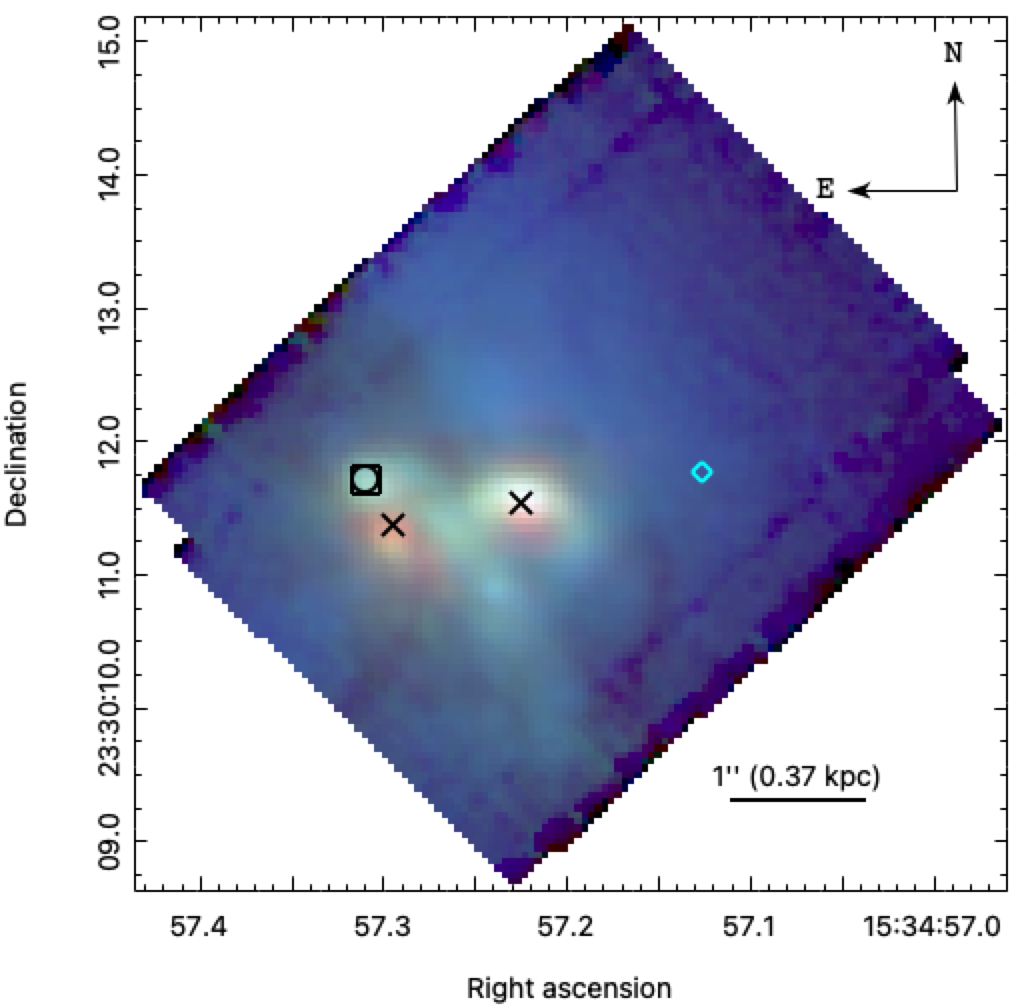}

\caption{ {\it Three-colour near-infrared image of Arp 220} obtained by combining NIRSpec narrow-band images tracing the continuum emission at $\sim 1\ \mu$m (blue), $3\mu$m (green) and $5\ \mu$m (red). The positions of the W and E nuclei are marked with {\sc x} symbols and match those of the nuclear pc-scale disks observed in the mm with ALMA (\citealt{Scoville2017}). The box-circle symbol marks the position of a bright cluster. The cyan diamond indicates a region with fast multi-phase outflows.}\label{fig:fig1}
\end{figure}

\subsection{Astrometric registration}

Before analysing the Arp 220 data-cubes, we tested whether they are aligned with each other. We extracted NIRSpec equivalent narrow-band images from the wavelength range $1.85-1.87\ \mu$m covered by both band1 and band2 cubes; these images are perfectly aligned in the pixel space, although a small discrepancy in the WCS coordinates registration of the reference pixel introduces an offset of $\delta$RA $=-0.0047\arcsec$ 
and $\delta$DEC $= -0.0141\arcsec$. 
Similarly, we obtained additional narrow-band images covering the range $3.00-3.02\ \mu$m from the band2 and band3 cubes; also in this case, the images are perfectly aligned in the pixel space, but present a discrepancy in WCS registrations, with $\delta$RA $=+0.0017\arcsec$ 
and $\delta$DEC $= 0.0071\arcsec$. 
Therefore, we corrected the WCS registrations of the band2 and band3 cubes to match the coordinates in band1. 

The absolute astrometric registration is performed using the ALMA 0.1\arcsec \ resolution maps of continuum emission at 2.6 mm presented by \citet[][project 2015.1.00113.S]{Scoville2017}, and the  0.03\arcsec \ resolution map of 1.3~mm continuum (project 2017.1.00042.S). 
We matched the central position of the two nuclei in the NIRSpec narrow-band image, obtained by collapsing the band3 cube in the range $5.1-5.12\mu$m, and in the ALMA images. The narrow wavelength range used to identify the nuclei in the NIR has been preferred to those at shorter wavelengths, as the latter are significantly more affected by dust extinction (see Sect.~\ref{sec:analysis}). 
The astrometry corrections we applied to the three NIRSpec cubes are $\delta$RA $=-0.3028\arcsec$ and $\delta$DEC $= -0.0209\arcsec$. This offset is consistent with the expected accuracy of JWST pointing without a dedicated target acquisition procedure (\citealt{Boker2023}).

\section{Analysis}\label{sec:analysis}

The two Arp 220 nuclei separated by $\sim 1$\arcsec\ are visible in the infrared (e.g. \citealt{Scoville1998}), millimetre (e.g. \citealt{Scoville2017, Lamperti2022}), radio (e.g. \citealt{Baan1995}), and X-ray (e.g. \citealt{Lockhart2015}). Figure \ref{fig:fig1} shows the composite near-infrared image of Arp~220 obtained from the combination of three NIRSpec narrow-band images extracted from the reduced data-cube at $\sim 1\mu$m (blue), $3\mu$m (green) and $5\mu$m (red), covering a Field of View (FoV) of $5\arcsec \times 4\arcsec$. The position of the W and E nuclei are marked for visual purposes.

The Arp~220 nuclear morphology is strongly influenced by dust extinction, even in the NIR (see also Fig. \ref{fig:figA1}). 
The continuum emission appears as a crescent arc above the W nucleus at $\sim 1\mu$m and $\sim 3\mu$m. The E nucleus is barely visible at these wavelengths, but a few bright, clumpy regions likely associated with clusters can be identified (see also \citealt{Scoville1998}). Among them, we marked with a box-circle the position of the brightest clump at the north of the E nucleus; hereinafter, we refer to it as the NE cluster, although it may be part of a more extended structure connecting the blue clumps in the regions between the nuclei in Fig. \ref{fig:fig1}. The emission at $\sim 5\ \mu$m is mostly dominated by the two unresolved nuclei, marked in the figure with two crosses.

\begin{figure*}[!htb]
\centering 
\includegraphics[width=\textwidth,trim= 20 0 40 0,clip]{{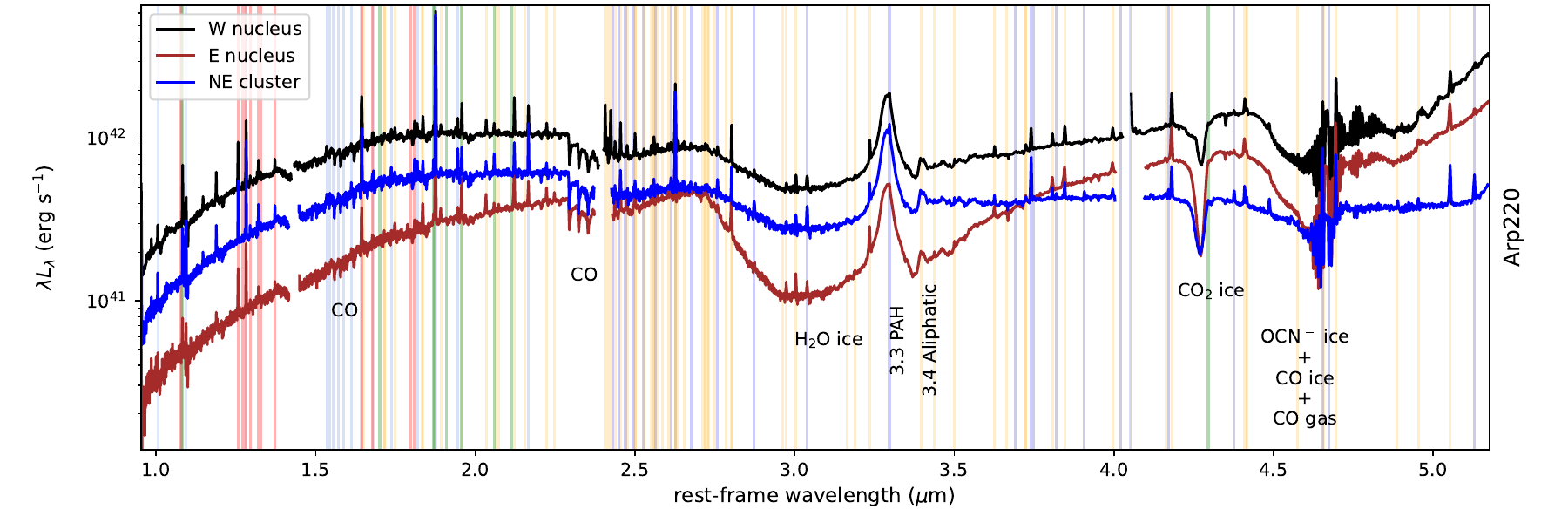}}

\caption{ {\it Arp 220 NE cluster, W and E nuclear spectra from the three NIRSpec gratings.} The spectra are shown in luminosity as a function of the rest frame wavelength. 
Vertical blue lines mark the positions of hydrogen transitions (dark to light blue colours indicate distinct hydrogen series); orange lines identify H$_2$ lines; red lines are associated with \feii transitions; green lines mark \hei emission features. Strong broad-band absorption and emission features are also labelled in the figure. }\label{fig:totalspectra}
\end{figure*}

\subsection{Spectral extraction}

We extracted the spectra of the two nuclei and the NE cluster shown in Fig. \ref{fig:fig1} for each of the three NIRSpec gratings. We used apertures of three spaxels in radius, corresponding to $0.15\arcsec$ or $ 55$ pc. This aperture size broadly corresponds to two NIRSpec spatial resolution elements at 5 $\mu$m (FWHM $\sim 0.15\arcsec$; \citealt{DEugenio2023}), thus ensuring that the entire point-source flux is captured at all wavelengths. 

The three spectra are reported in Fig. \ref{fig:totalspectra}; they are shown in rest-frame, taking into account their relative redshifts: $z_W = 0.01774$, $z_E = 0.01839$, and $z_{NE} = 0.01870$ (see Table \ref{tab:properties}). These values are obtained by measuring the centroid of the brightest hydrogen lines, and are comparable with those inferred from sub-mm/mm ALMA spectroscopy at angular resolutions similar to those of NIRSpec (see Sect. \ref{sec:zspec}).  

The spectra show a few gaps in wavelength coverage
in the middle of the three bands, due to the gap between the two NIRSpec detectors. The overlapping wavelength ranges at $\sim 1.85\ \mu$m and $\sim 3\ \mu$m show a perfect match between the different bands, demonstrating a consistent flux calibration. We note that no aperture corrections are taken into account in this work.

\begin{table*}[h]
\centering
\caption{Properties of Arp220 nuclear regions}%
\begin{threeparttable}
\begin{tabular}{lccc}
\hline
  & W nucleus & E nucleus & NE cluster\\
\hline
RA (J2000) & 15:34:57.224  & 15:34:57.294  & 15:34:57.309 \\
DEC (J2000) & +23:30:11.515 & +23:30:11.353 & +23:30:11.698 \\
redshift (\hi) & 0.01774 $\pm$ 0.00001 & 0.018396 $\pm$ 0.00001 &  0.018704 $\pm$ 0.00001\\
redshift (stars) & $0.0179\pm 0.0003$ & $0.0183\pm 0.0002$ & $0.0186\pm0.0001$\\
SFR(radio/far-IR) [\Msunyr] & 110 -- 150 & 60 -- 80 & --\\
SFR(\Paa$_{corr}$) [\Msunyr] & $0.52_{-0.06}^{+0.16}$ & $0.17\pm 0.02$ & $0.94\pm 0.02$\\
SFR(PAH $3.3\ \mu$m) [\Msunyr] & $3.4\pm 0.2$& $1.5\pm 0.2$ & $1.9\pm 0.2$ \\
A$_v$ [mag]  & $10.6_{-0.5}^{+1.7}$ & $14.1_{-0.3}^{+0.6}$ & $11.2\pm 0.1$\\
log ($n_e$/[cm$^{-3}$] & $3.7\pm 0.1$ & $4.5\pm 0.4$ & $3.6\pm 0.1$ \\
\hline
\end{tabular} 
\begin{tablenotes}[para,flushleft]
Notes: SFR measurements are corrected for dust extinction using the A$_V$ reported in the table; the latter are obtained from Eq. \ref{eq:EBVBr}, assuming $R_V = 3.1$. Radio and far-IR based SFR measurements come from \citet{Varenius2016} and \citet{Pereira2021}, respectively.  
  \end{tablenotes}
  \end{threeparttable}
\label{tab:properties}
\end{table*}

\subsection{Identification of the lines}


The identification of the multitude of emission lines detected in NIRSpec spectra of Arp 220, and marked in Fig. \ref{fig:totalspectra} with vertical lines, was performed using compilations of NIR features in star-forming and active galaxies from the literature (e.g. \citealt{Lutz2000, Koo2016, Lamperti2017, Lee2017, Riffel2019}). We also use the NIST atomic spectra database\footnote{\url{https://www.nist.gov/pml/atomic-spectra-database}} (\citealt{Ralchenko2005}) and the HITRAN (high-resolution transmission molecular absorption database\footnote{\url{https://hitran.org/}}, \citealt{Rothman2021}) to provide probable identifications of faint lines that, to our knowledge, have not previously been reported in the literature.

Figures \ref{fig:portion1}-\ref{fig:portion6} report all species identified in the spectra of the Arp 220 nuclei and the NE cluster, in comparison with those in the nuclear regions of another nearby interacting LIRG observed with NIRSpec, VV114 (e.g. \citealt{Rich2023}). Firmly detected line transitions are indicated with solid vertical lines, and are also reported in Table \ref{tab:linelist}. Probable identifications are marked in the figures in the Appendix with dotted vertical lines. All emission lines are well visible in the spectra of the south-west nucleus of VV114 (named SWc2 by \citealt{GonzalezAlfonso2023}), at higher signal-to-noise and with narrower profiles than in Arp 220.
All three nuclear regions of VV114, namely NE, SW, and SWc2, have been suggested to harbor deeply obscured AGN (e.g. \citealt{Rich2023, GonzalezAlfonso2023}). Hence, they serve as an optimal benchmark for comparison with the nuclei of Arp 220, which are also presumed to contain obscured AGN.


\begin{table*}[h]
\centering
\caption{Arp 220 emission line list in the W and E nuclei, and NE cluster (band1 spectra) }%
\begin{threeparttable}
\begin{tabular}{|lc|c|c|c|}
\hline

\hline
line & $\lambda_{vac}$ & $f_W$ &  $f_E$ & $f_{NE}$ \\
     &  ($\mu$m) &  ($10^{-18}$ \ergs cm$^{-2}$)  &  ($10^{-18}$ \ergs cm$^{-2}$) &  ($10^{-18}$ \ergs cm$^{-2}$)  \\
\hline

\siii & 0.953321 & 598.0$_{-0.4}^{+0.4}$ & 54.9$_{-1.2}^{+1.1}$ & 438.0$_{-1.0}^{+1.8}$ \\
\Pae & 0.954865 & 87.9$_{-1.0}^{+1.5}$ & 7.8$_{-0.8}^{+1.0}$ & 41.7$_{-0.8}^{+1.9}$ \\
\ci & 0.982412 & 11.3$_{-1.1}^{+1.2}$ & 3.3$_{-0.6}^{+0.8}$ & 8.8$_{-0.9}^{+1.8}$ \\
\ci & 0.985295 & 63.7$_{-0.5}^{+1.2}$ & 15.5$_{-0.9}^{+1.4}$ & 28.8$_{-0.3}^{+0.0}$ \\
\Pad & 1.005219 & 112.0$_{-1.5}^{+1.4}$ & 15.8$_{-1.0}^{+1.2}$ & 67.6$_{-0.8}^{+1.6}$ \\
\sii & 1.028955 & 48.3$_{-2.7}^{+5.2}$ & 5.7$_{-1.5}^{+4.2}$ & 11.6$_{-1.6}^{+2.5}$ \\
\sii & 1.032332 & 62.6$_{-2.2}^{+2.5}$ & 0.9$_{-0.0}^{+0.0}$ & 13.6$_{-1.6}^{+2.7}$ \\
\sii & 1.033924 & 63.1$_{-1.0}^{+2.6}$ & 5.3$_{-1.9}^{+2.4}$ & 17.2$_{-1.4}^{+2.6}$ \\
\sii & 1.037334 & 39.3$_{-0.3}^{+0.1}$ & 0.9$_{-0.0}^{+4.4}$ & 11.1$_{-2.7}^{+6.1}$ \\
$[$N\,{\sc{i}}] & 1.040059 & 42.8$_{-0.0}^{+0.1}$ & 4.2$_{-3.2}^{+2.9}$ & 3.8$_{-2.8}^{+2.5}$ \\
\hei & 1.083331 & 628.0$_{-1.1}^{+2.2}$ & 41.2$_{-0.4}^{+1.3}$ & 185.0$_{-0.7}^{+1.0}$ \\
\Pag & 1.094116 & 167.0$_{-1.6}^{+1.7}$ & 25.5$_{-0.6}^{+1.1}$ & 205.0$_{-1.9}^{+1.7}$ \\
\pii & 1.147134 & 101.0$_{-0.9}^{+1.4}$ & 9.4$_{-0.5}^{+1.0}$ & 45.7$_{-0.7}^{+1.2}$ \\
\pii & 1.188605 & 328.0$_{-0.7}^{+1.3}$ & 27.8$_{-0.3}^{+0.7}$ & 126.0$_{-0.8}^{+0.9}$ \\

$[$Ni\,{\sc{ii}}] & 1.191000 & 31.5$_{-0.6}^{+1.3}$ & 5.0$_{-0.7}^{+0.6}$ & 19.6$_{-0.8}^{+1.1}$ \\
\feii & 1.257021 & 711.0$_{-4.3}^{+10.6}$ & 78.0$_{-1.8}^{+3.6}$ & 371.0$_{-3.8}^{+7.2}$ \\
\feii & 1.270690 & 44.5$_{-2.2}^{+5.1}$ & 0.8$_{-0.0}^{+1.7}$ & 10.8$_{-3.6}^{+6.6}$ \\
\feii & 1.278789 & 75.6$_{-1.6}^{+3.7}$ & 3.3$_{-2.5}^{+6.0}$ & 26.7$_{-3.1}^{+5.5}$ \\
\Pab & 1.282167 & 1170.0$_{-7.0}^{+6.7}$ & 124.0$_{-2.6}^{+4.6}$ & 822.0$_{-4.1}^{+6.5}$ \\
\feii & 1.294453 & 71.7$_{-2.5}^{+6.6}$ & 7.1$_{-2.0}^{+2.8}$ & 24.6$_{-2.0}^{+5.3}$ \\
\feii & 1.320911 & 289.0$_{-4.5}^{+7.0}$ & 26.9$_{-2.0}^{+3.2}$ & 130.0$_{-3.8}^{+4.6}$ \\
$[$O I] & 1.316871 & 94.2$_{-2.5}^{+5.3}$ & 4.0$_{-2.4}^{+3.0}$ & 22.6$_{-2.2}^{+5.3}$ \\
\feii & 1.372200 & 185.0$_{-4.3}^{+6.1}$ & 25.0$_{-1.2}^{+2.9}$ & 102.0$_{-2.0}^{+4.3}$ \\
Br 13 & 1.611378 & 51.6$_{-3.6}^{+6.9}$ & 9.2$_{-3.0}^{+3.5}$ & 54.0$_{-3.8}^{+5.2}$ \\
Br 12 & 1.641174 & 71.3$_{-4.2}^{+6.6}$ & 11.0$_{-4.5}^{+2.9}$ & 7.6$_{-3.8}^{+4.0}$ \\
\feii & 1.644002 & 1340.0$_{-5.7}^{+11.1}$ & 209.0$_{-3.9}^{+8.5}$ & 787.0$_{-8.6}^{+9.6}$ \\
\feii & 1.677300 & 160.0$_{-5.2}^{+6.6}$ & 20.3$_{-2.7}^{+6.4}$ & 67.3$_{-5.1}^{+5.8}$ \\
Br 11 & 1.681118 & 166.0$_{-7.0}^{+7.2}$ & 27.7$_{-2.5}^{+5.1}$ & 117.0$_{-3.3}^{+5.5}$ \\
H$_2$ & 1.714881 & 14.9$_{-11.3}^{+72.8}$ & 2.6$_{-0.1}^{+40.5}$ & 2.5$_{-0.0}^{+0.0}$ \\
H$_2$ & 1.714958 & 88.8$_{-59.5}^{+4.6}$ & 30.3$_{-20.1}^{+12.3}$ & 29.4$_{-7.0}^{+5.0}$ \\

H$_2$ & 1.748101 & 176.0$_{-8.6}^{+5.6}$ & 43.8$_{-8.3}^{+18.9}$ & 66.1$_{-5.9}^{+1.2}$ \\

\Brten & 1.736692 & 202.0$_{-6.4}^{+6.3}$ & 17.4$_{-2.5}^{+4.5}$ & 111.0$_{-7.2}^{+6.9}$ \\
\Bre & 1.817916 & 245.0$_{-5.7}^{+11.5}$ & 34.0$_{-4.5}^{+8.3}$ & 168.0$_{-4.5}^{+12.7}$ \\
H$_2$ & 1.834123 & 23.1$_{-4.2}^{+11.9}$ & 57.8$_{-22.0}^{+12.6}$ & 37.6$_{-7.0}^{+8.2}$ \\
H$_2$ & 1.835913 & 370.0$_{-6.9}^{+12.4}$ & 109.0$_{-13.4}^{+39.8}$ & 145.0$_{-3.7}^{+10.8}$ \\

\hline\hline

\hline
\end{tabular} 
\begin{tablenotes}[para,flushleft]
Notes: This list comprises all emission lines detected in band1; see Tables \ref{tab:linelist2} and \ref{tab:linelist3} for the ones in band2 and band3, respectively. The reported fluxes are obtained within apertures of 0.3\arcsec \ in diameter from single Gaussian fits, without dust- and aperture-corrections.   
  \end{tablenotes}
  \end{threeparttable}
\label{tab:linelist}
\end{table*}


\subsection{Modelling of the continuum emission}\label{sec:continuum}

To fit the stellar continuum, we used the penalised-pixel-fitting routine pPXF (\citealt{Cappellari2017}). We used the MARCS synthetic library of stellar spectra to model the stellar continuum (\citealt{Gustafsson2008}); these templates have a constant spectral resolution $\Delta\lambda/\lambda$ = 20\,000 ($\sigma \sim 6$ \kms) and cover the range $0.13-20\ \mu$m with wavelengths in vacuum. Although not as reliable as an empirical stellar library, MARCS templates were preferred to E-MILES (\citealt{Vazdekis2016}), which have lower spectral resolution than our NIRSpec data, and XSL (\citealt{Verro2022}), which have narrower wavelength coverage and are affected by telluric features.

We also adopted a sixth-order multiplicative polynomial to better reproduce the spectral continuum shape. 
In addition, we used a set of gas emission line templates (parametrised with Gaussian profiles) together with the stellar ones to guide the fit. We conducted separate fits for the band1 ($0.95-1.86 \mu$m) and band2 ($1.63-3.11\ \mu$m) cubes, with the latter restricted to wavelengths below 2.4 $\mu$m, corresponding to the cutoff of the detector gap. For band1, the wavelengths falling within the NIRSpec detector gap were appropriately masked during the fit procedure.

Since no prominent stellar features are observed at wavelengths above $2.4\ \mu$m, and because of the presence of dust emission and ice absorption (e.g. \citealt{Donnan2024}), we preferred to model the pseudo-continuum with a Savitzky–Golay filter after masking all prominent emission and absorption features. 

In Figs. \ref{fig:portion1}-\ref{fig:portion3} we report the reconstructed pPXF best-fit continuum emission, while Figs. \ref{fig:portion4}-\ref{fig:portion6} show the pseudo-continuum shape in the two Arp220 nuclei and the NE cluster. We used the (pseudo-)continuum-subtracted spectra to model the emission lines, and hence derive the ISM kinematic and physical properties in these three regions.

\subsection{Modelling of the emission lines}

After subtracting the (pseudo-)continuum, we modelled the most relevant gas emission lines using a combination of Gaussian profiles. In particular, we modelled a set of \hi recombination lines - Paschen (Pa), Brackett (Br), Pfund (Pf), and Humphreys (Hu) series - and molecular hydrogen features, as well as singly ionised iron transitions. 

The sets of \hi, H$_2$ and \feii transitions are modelled using independent fits, to account for the fact that  these three species can trace different gas phases and can have therefore different systemic velocities and line widths (e.g. \citealt{May2017}). 
All transitions were fitted using  the Levenberg–Markwardt least-squares fitting code CAP-MPFIT (\citealt{Cappellari2017}). 

The transitions of a given species were fitted simultaneously (e.g. \citealt{Perna2015a}), as it allows to better constrain the properties of the transitions associated with lower S/N. Namely, we constrained the wavelength separation between emission lines in accordance with atomic physics (considering their vacuum wavelengths); moreover, we
fixed their widths (in \kms) to be the same for all the emission lines. Some emission lines are covered by two cubes (e.g. \Bre at $\sim 1.82\ \mu$m, in band1 and band2 spectra); for these lines we obtained an average profile by considering the errors from individual cubes as weights. 

For each species we performed two fits, with one and two Gaussian components per emission line respectively. The number of Gaussian sets (i.e. distinct kinematic components) used to model the lines of a specific species was determined on the basis of the Bayesian information criterion (BIC; \citealt{Schwarz1978}). 

The measured fluxes of all identified emission line transitions are reported in Tables \ref{tab:linelist}, \ref{tab:linelist2} and \ref{tab:linelist3}. To streamline the presentation, all tabulated fluxes are derived from single Gaussian fits; detailed Gaussian decomposition of bright line profiles is discussed in Sect. \ref{sec:results}.

\section{Results}\label{sec:results}

In this section, we report the general characterisation of the nuclei and the NE cluster, in terms of spectral shape, extinction, electron density, and presence of outflows. These properties are discussed in Sect. \ref{sec:discussion} to investigate the possible presence of AGN signatures. 


\subsection{Systemic redshift of the two nuclei}\label{sec:zspec}

The comparison of the three selected regions in Arp 220 was performed after reporting the individual spectra in luminosity ($\lambda L_\lambda$) as a function  of the rest-frame wavelength in vacuum. 
We used the centroid of the \hi lines to determine the redshifts for the three spectra. In fact, these lines are quite symmetric and narrow and, therefore, are likely less affected by outflows and perturbed motions than other lines (see \citealt{Perna2020,Wheeler2020}). In contrast, the brightest H$_2$ and \feii lines show prominent blue wings and are blue-shifted by a few 10s \kms with respect to the \hi lines (Table \ref{tab:kinematics}).

The redshifts inferred from the atomic hydrogen lines are reported in Table \ref{tab:properties}. These values are consistent with those obtained from the stellar continuum modelling, within the uncertainties (see Table \ref{tab:properties}).  Moreover, our measurements match the ones obtained from the study of sub-mm/mm molecular line transitions 
as reported in \cite{Sakamoto2021b}, $z_W \sim 0.01768$ and $z_E \sim 0.01808$, and \citet{Pereira2021}, $z_W \sim 0.01765$ and $z_W \sim 0.01797$. In particular, for the W nucleus, all measurements are consistent within $\sim 20$ \kms. The molecular lines in the E nucleus are quite asymmetric (see Fig. 1 in \citealt{Sakamoto2021b}), resulting in slightly larger discrepancies compared to our NIR redshift. However, all measurements remain consistent within $\lesssim 100$ \kms, suggesting minimal relative motions between the various components. 


\subsection{Qualitative comparison between Arp 220 nuclear regions and the bright cluster}


The spectra of the two nuclei and the NE cluster contain a high number of emission lines: among them hydrogen lines (from the Pa, Br, Pf, and Hu series), as well as helium, iron, and molecular hydrogen transitions. Notably, \hi lines appear a few times fainter in the E spectrum compared to the W and NE spectra, likely due to higher extinction levels (see Sect. \ref{sec:extinction}) and lower SFR (Sect. \ref{sec:SFR}). The W nucleus displays the most intense H$_2$ lines, approximately $2\times$ stronger than those in the E and NE spectra.

At 3.3 $\mu$m, all three spectra show a broad emission feature produced by polycyclic aromatic hydrocarbons (PAHs); the 3.4 $\mu$m aliphatic features are also present in all three locations. PAHs and aliphatic transitions are stronger in the W nucleus, by a factor of $\sim 3$ (2) with respect to the E nucleus (NE cluster; see Sect. \ref{sec:SFR}). This suggests that the SFR is higher in the W nucleus (as PAHs trace recent star formation activity; Sect. \ref{sec:SFR}). 


The prominent CO absorption bands at $\sim 1.6\  \mu$m and between 2.3 $\mu$m and 2.4 $\mu$m, produced in the atmospheres of giant and supergiant stars (\citealt{Oliva1995}) 
are clearly visible in the two nuclei as well as in the NE cluster.

Various other broad-band absorption features are visible as well, mostly produced by solid state molecules of water-ice (H$_2$O, at 3 and 4.5 $\mu$m), and carbon monoxide (CO, at 4.67 $\mu$m) transitions. These absorption features are stronger in the nuclear regions than in the nearby cluster; the $3.3-3.8 \mu$m ``ice band wing'' (\citealt{Gibb2004}) is observed only in the two nuclei. The strength of these absorption bands is therefore likely correlated with the abundance of dust, given that the two nuclear regions are the most obscured in the mm range (\citealt{Sakamoto2021a}). Conversely, the carbon dioxide ice transitions $^{12}$CO$_2$ at 4.27  $\mu$m and $^{13}$CO$_2$ at 4.38  $\mu$m  have rather peculiar shapes in the three spectra: the former is stronger in the E nucleus, but more asymmetric in the W nucleus (with a prominent blue shoulder), while the NE cluster shows intermediate strength with respect to the nuclei. The $^{13}$CO$_2$ is deeper and narrower in the W nucleus.  

The W and E nuclei of Arp 220 show $^{12}$CO ro-vibrational lines up to J$_{low}$ = 23 (Fig. \ref{fig:portion6}), indicating the presence of high excitation gas. Such transitions could provide a further indirect evidence of AGN activity. Detailed modelling is required to infer the CO gas properties, and hence test the more likely mechanisms responsible for such high excitation (e.g. \citealt{Baba2022, GonzalezAlfonso2023, Buiten2023, GarciaBernete2024, Pereira2024a}). This analysis will be presented in a forthcoming paper (Buiten, van der Werf et al., in prep.).


Differences in the three spectra can also be appreciated from a qualitative comparison between their continuum shapes. Going from the shortest wavelengths to $\sim 1.8\ \mu$m, all three sources show a pronounced steepening in their spectra, likely dominated by the diminishing effect of dust obscuration with increasing wavelength (see also \citealt{Engel2011}). In the range $1.8-2.3\ \mu$m the W and NE spectra flatten, while the E spectrum still shows the same steepening observed at shortest wavelengths; this could suggest an higher extinction at the position of the E nucleus, due to the dust lane visible in reddish colours in Fig. \ref{fig:fig1}. Between 2.3 and 5$\mu$m, the nuclear regions show a mild increase in luminosity while NE is almost constant. At $\sim 5\mu$m, the spectrum of the NE cluster reveals a modest rise in luminosity, while the two nuclei exhibit a much more pronounced increase likely due to the presence of warm dust (\citealt{Armus2007}). It is possible therefore that dust-absorbed light is re-emitted at wavelengths $\gtrsim 2\ \mu$m in the two nuclei; the higher dilution of CO bands at $\sim 2.3 - 2.4\ \mu$m in the nuclei compared to the NE cluster (Fig. \ref{fig:portion3}) also supports this possibility, as suggested by \citealt{Engel2011}.

In the next sections, we will briefly discuss the general properties of some of these components. 
We do not pursue the full spectral energy distribution
modelling further in this work, but instead will use a subsequent paper to focus on a more detailed investigation.





\subsection{Extinction}\label{sec:extinction}

Extinction can have an important effect on the estimated source luminosity and SFR, and we therefore consider different methods for deriving its value. We compared pairs of emission lines at wide wavelength separation, for which the intrinsic ratio of line strengths is known. We calculated the colour excess in terms of \Pab and \Brg, 

\begin{equation}\label{eq:EBVBr}
\begin{split}
E(B-V) & = \frac{2.5}{k(Pa\beta) - k(Br\gamma)} \ log_{10} \left( \frac{\left(Pa\beta/Br\gamma\right)_{obs}}{\left(Pa\beta/Br\gamma\right)_{int}} \right)\\ & = 5.22 \ log_{10} \left( \frac{\left(Pa\beta/Br\gamma\right)_{obs}}{5.88} \right)
\end{split}
\end{equation}

where the intrinsic ratio \Pab/\Brg is set to 5.88 assuming an electron temperature and a density of 10$^4$ K and 10$^3$ cm$^{-3}$ respectively (e.g. \citealt{Osterbrock2006}). We also used the \Pfg and \Brd to obtain an independent estimate for the colour excess

\begin{equation}\label{eq:EBVPf}
\begin{split}
E(B-V) & = \frac{2.5}{k(Br\delta) - k(Pf\gamma)} \ log_{10} \left( \frac{\left(Br\delta/Pf\gamma\right)_{obs}}{\left(Br\delta/Pf\gamma\right)_{int}} \right)\\ & = 8.95 \ log_{10} \left( \frac{\left(Br\delta/Pf\gamma\right)_{obs}}{1.71} \right)
\end{split}
\end{equation}

where the intrinsic ratio \Brd/\Pfg is set to 1.71 (with the same assumptions as reported above). The coefficients in Eqs. \ref{eq:EBVBr} and \ref{eq:EBVPf} were derived assuming a \cite{Cardelli1989} extinction law. 


We derived an additional estimate of the colour excess from the flux ratio of \feii at 1.257 and 1.644 $\mu$m (e.g. \citealt{Riffel2014}), which are the two strongest iron lines in the NIR regime,

\begin{equation}\label{eq:EBVfe}
E(B-V) = 8.95 \ log_{10} \left( \frac{1.34}{\left([Fe II]_{1.257}/[Fe II]_{1.644}\right)_{obs}} \right)
\end{equation}

where 1.34 is the intrinsic ratio (we note, however, that this is still a matter of some debate, and different values are reported in the literature, from 1.04 to 1.49, see \citealt{Eriksen2009}). 

We measured a colour excess of the order of a few magnitudes ($\sim 3-5$) in the three regions of interest, corresponding to A$_V \sim 10-14$ (assuming R$_V = 3.1$). 
We do not observe significant differences in the measurements obtained from the three diagnostic ratios above mentioned (Eqs. \ref{eq:EBVBr}, \ref{eq:EBVPf}, \ref{eq:EBVfe}). Therefore, hereinafter we will refer to the results obtained from Eq. \ref{eq:EBVBr}. The most extreme extinction is associated with the E nucleus, with A$_V \sim 14$, consistent with the presence of the strongest steepening in continuum emission at wavelengths $\lesssim 2\ \mu$m with respect to the W nucleus and the NE cluster (with A$_V \sim 11$, see Table \ref{tab:properties}).

In previous studies, an extinction of $\approx 6$ mag in the innermost nuclear regions of Arp 220 was reported, employing diagnostics such as the \hi flux ratios \ha/\hb (\citealt{Perna2020}), \ha/\Pab (\citealt{Gimenez2022}), and \Paa/\Brg (\citealt{Engel2011}). More extreme measurements were obtained from mid-infrared diagnostics: for instance, \citet{Haas2001} reported an extinction of a few tens of magnitudes. 
This suggests that NIRSpec observations may only probe the outer gaseous layers, and that the total extinction towards the core of these dusty regions can be much higher than $A_V \sim 10-14$.




\subsection{Electron density}\label{sec:density}

We derived the electron density in the line-emitting regions surrounding the E, W, and NE regions of Arp 220 via the comparison of observed line strength ratios of \feii transitions with theoretical values. 

We used the ratio \feii 1.677 $\mu$m/1.644 $\mu$m to infer electron densities in the range log \Ne \ $\approx 3.5- 4.5$ cm$^{-3}$ for the three regions of interest (from Eq. 6 in \citealt{Koo2016}). For the two nuclei, for which two kinematic components are required to fit the total profiles, we consider the narrow components (as the broad one is undetected in the faint \feii1.677 $\mu$m). These measurements, reported in Table \ref{tab:properties}, are discussed in the context of previous measurements in Sect. \ref{sec:density_discussion}.


\begin{figure}[!htb]
\centering 
\includegraphics[width=0.5\textwidth,trim= 20 0 0 0,clip]{{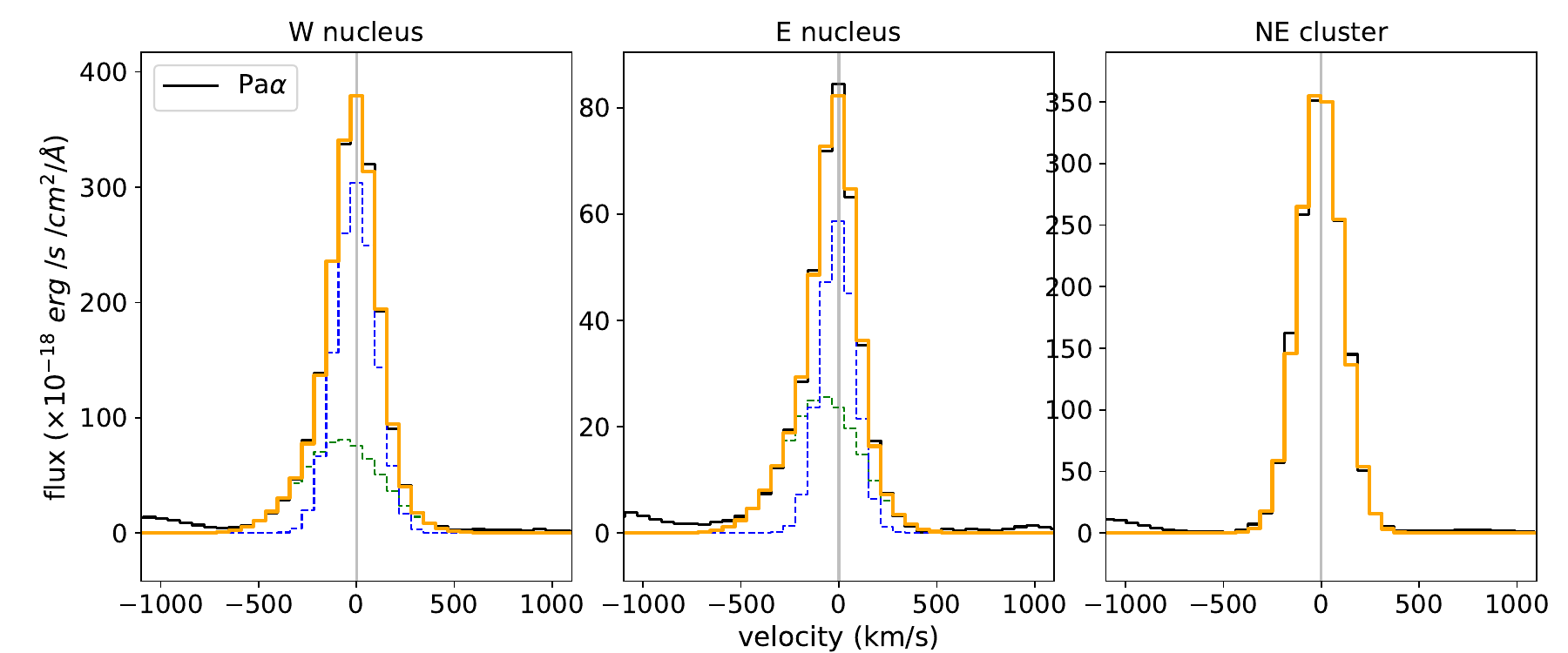}}
\includegraphics[width=0.5\textwidth,trim= 20 0 0 24,clip]{{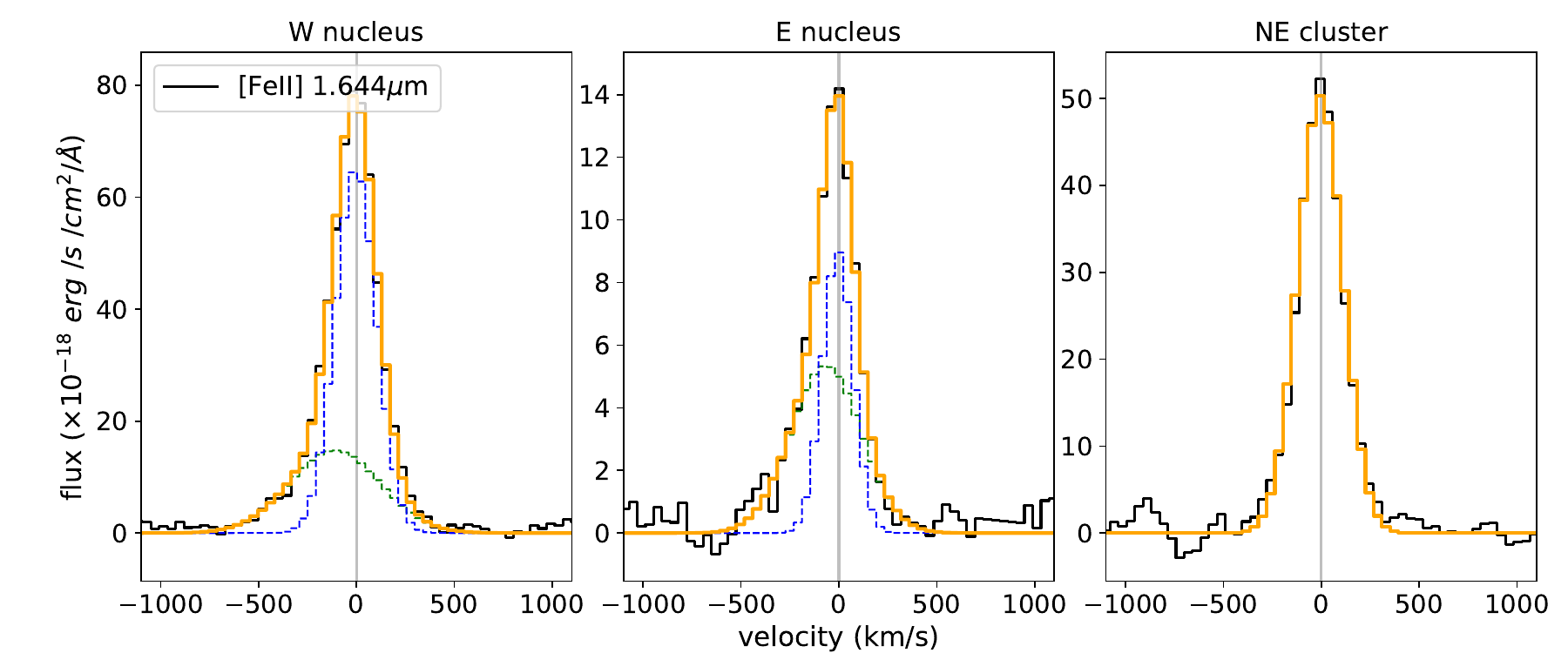}}
\includegraphics[width=0.5\textwidth,trim= 20 0 0 24,clip]{{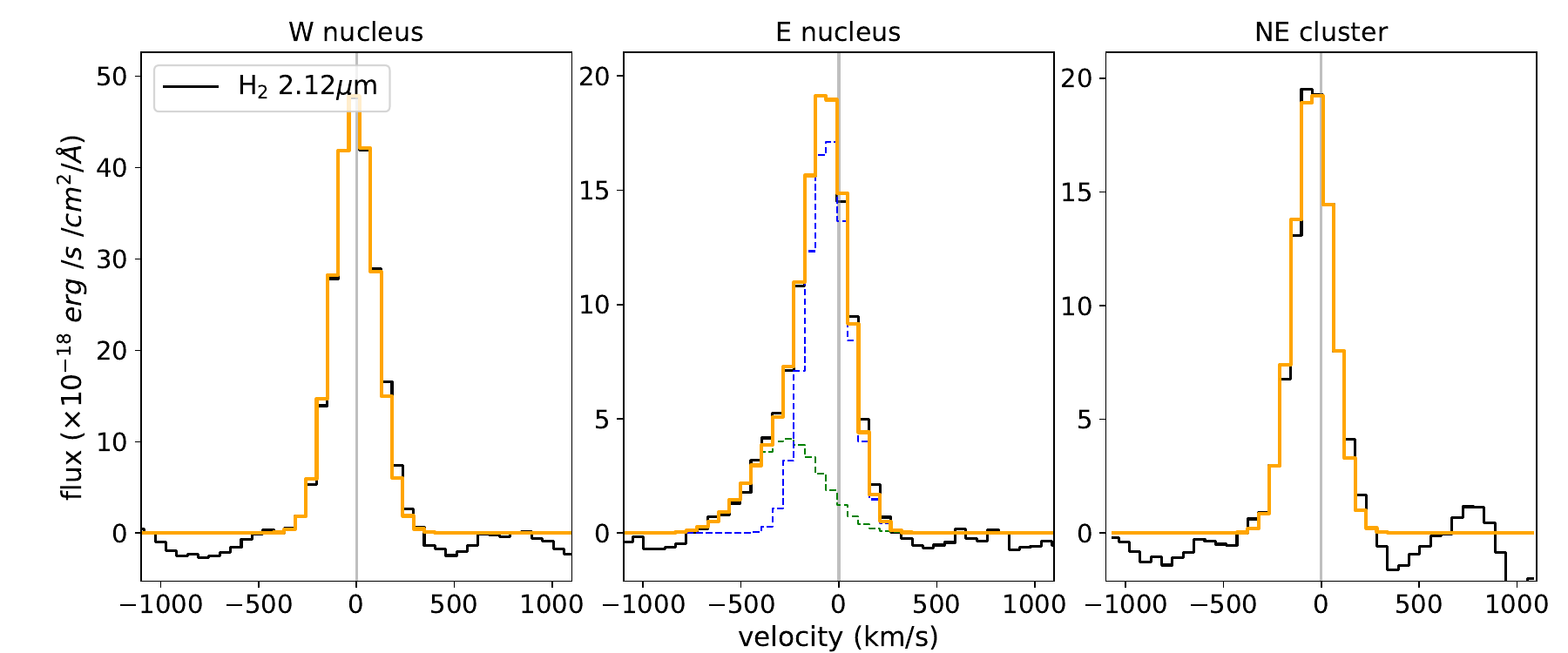}}

\caption{ {\it Continuum-subtracted integrated spectra of the three regions of interest in the vicinity of the \Paa, \feii 1.644 $\mu$m, and H$_2$ 2.12 $\mu$m.} 
The original spectra (in black) are reported in velocity space. The best-fit emission line profiles are shown in orange; the individual Gaussian profiles used to reproduce the line shapes are reported in blue and green. Broad and asymmetric profiles are found in the nuclear spectra. The vertical lines mark the zero-velocity as inferred from the narrow component of \hi lines.}\label{fig:outflows}
\end{figure}







\subsection{Nuclear outflows}\label{sec:outflow}

A distinctive feature of Arp 220's nuclear environment is the presence of powerful and complex outflows. Compact and collimated cold molecular outflows, with velocities of a few 100s \kms, a bipolar morphology and an extension of $\sim 120$ pc, are observed in both nuclei (e.g. \citealt{Wheeler2020}). Broader line profiles are instead observed in the ionized and atomic gas: for instance, \citet{Perna2021} reported $W80$ velocities (defined as the difference between the 90th- and 10th-percentile velocities of the fitted line profile) of $\sim 800$ \kms. However, these measurements have been obtained from observations at lower spatial resolution ($\sim 200$ pc), and likely cannot be used to properly resolve the outflow kinematics on smaller scales. 

Figure \ref{fig:outflows} shows the integrated spectra in the vicinity of the \Paa (top), \feii1.644 $\mu$m (centre), and H$_2$ 2.12$\mu$m (bottom) lines for the three regions of interest. The NE cluster shows symmetric and narrow profiles; conversely, broad and asymmetric (preferentially blue-shifted) line shapes are observed in both nuclear regions, with outflowing gas at velocities up to $-500$ \kms. In this work, we consider as outflow velocity $v_{max} = \Delta V_{broad} + 2\sigma_{broad}$, where $\sigma_{broad}$ is corrected for instrumental resolution (e.g. \citealt{Fluetsch2021}). In Table \ref{tab:kinematics} we report the best-fit kinematic parameters, considering both non-parametric (e.g. $W80$), and parametric outflow velocity measurements (e.g. $v_{max}$; see caption of Table \ref{tab:kinematics} for details).

 A more detailed discussion about the presence of AGN winds possibly responsible for the broad and asymmetric profiles observed in the Arp 220 nuclei is presented in Sect. \ref{sec:outflow_discussion}. 

\begin{table}[h]
\tabcolsep 3.8pt 
\centering
\caption{Kinematic parameters}%
\begin{threeparttable}
\begin{tabular}{lcccc}

\multicolumn{5} {c} {W nucleus} \\
\hline
   & \hi & H$_2$ & \feii & \mgiv \\
\hline
$\Delta V_1$ & $0\pm 1$ & $-1_{-4}^{+26}$ & $0_{-1}^{+2}$ & $-111\pm 3$ \\
FWHM$_1$     & $238\pm 2$ & $250_{-6}^{+24}$ & $242_{-4}^{+6}$ & $234_{-10}^{+6}$\\
$\Delta V_2$ & $-72\pm 2$ & -- & $-110_{-8}^{+16}$ & --\\
FWHM$_2$ & $480_{-4}^{+6}$ & -- & $525_{-9}^{+24}$ & --\\
$W80$ & $370 \pm 3$ & $270\pm 20$ & $380\pm 3$ & $260\pm 10$\\
$V10$ & $-250 \pm 2$ & $-180\pm 15$ & $-270\pm 6$ & $-265\pm 10$\\
$V_{max}$ & $-480_{-4}^{+9}$ & $-220_{-3}^{+30}$ & $-560\pm 25$ & $-309\pm 8$\\
\hline\hline

\\
\multicolumn{5} {c} {E nucleus} \\
\hline
   & \hi & H$_2$ & \feii & \mgiv \\
\hline
$\Delta V_1$ & $0_{-2}^{+10}$ & $-40\pm 5$ & $-8_{-3}^{+8}$ & $-130_{-8}^{+14}$ \\
FWHM$_1$ & $205_{-27}^{+9}$ & $230 \pm 3$ & $180_{-70}^{+45}$ & $370_{-14}^{+20}$\\
$\Delta V_2$ & $-70_{-10}^{+60}$ & $-201_{-7}^{+13}$ & $-60\pm 30$ & --\\
FWHM$_2$ & $425_{-60}^{+50}$ & $405\pm 10$ & $420_{-60}^{+120}$ & --\\
$W80$ & $370 \pm 30$ & $440\pm 50$ & $380\pm 42$ & $400\pm 30$\\
$V10$ & $-250 \pm 20$ & $-370\pm 30$ & $-250 \pm 42$ & $-370\pm 20$\\
$V_{max}$ & $-430_{-45}^{+70}$ & $-540_{-10}^{+20}$ & $-430_{-90}^{+140}$ & $ -440\pm 20$\\
\hline\hline

\\
\multicolumn{5} {c} {NE cluster} \\
\hline
   & \hi & H$_2$ & \feii & \mgiv \\
\hline
$\Delta V_1$ &  $0_{-3}^{+4}$ & $-40\pm 5$ & $-14_{-8}^{+7}$ & $-81_{-3}^{+6}$\\
FWHM$_1$ & $260_{-3}^{+30}$ & $225_{-60}^{+10}$ & $210\pm 30$ & $260\pm 10$\\
$\Delta V_2$ & -- & -- & $11_{-6}^{+33}$ & --\\
FWHM$_2$  &  -- & -- & $340_{-40}^{+170}$ & --\\
$W80$ & $250\pm 10$ & $270_{-50}^{+10}$ & $290\pm 40$ & $305\pm 40$\\
$V10$ & $-160\pm 10$ & $-180_{-30}^{+10}$ & $-170\pm 30$ & $-240\pm 30$ \\
$V_{max}$ & $-230_{-3}^{+25}$ & $-230_{-60}^{+10}$ & $-190\pm 20$ & $-300\pm 6$\\
\hline\hline

\\
\multicolumn{5} {c} {NW outflow} \\
\hline
   & \hi & H$_2$ & \feii & \mgiv \\
\hline
$\Delta V_1$ & $-175_{-10}^{+30}$  & $-180_{-7}^{+16}$ & $-290_{-10}^{+15}$ &-\\
FWHM$_1$ & $130_{-20}^{+80}$ & $80_{-45}^{+20}$ & $640_{-30}^{+50}$ &-\\

$\Delta V_2$ & $-340_{-30}^{+40}$ & $-260_{-30}^{+70}$ & -- &-- \\
FWHM$_2$ & $650_{-50}^{+105}$ & $650_{-70}^{+60}$ &-- &-- \\
$\Delta V_3$ & -- & $85_{-6}^{+11}$ &-- &-- \\
FWHM$_3$ & -- & $210_{-30}^{+50}$ &-- &-- \\

$W80$ & $680_{-30}^{+90}$& $715\pm 10$  & $715\pm 40$ &--\\
$v10$ & $-690\pm 30$& $-590\pm 15$  &  $-670 \pm 35$ &--\\
$v_{max}$ & $-890_{-45}^{+130}$ & $-820\pm 35$  &  $-810_{-45}^{+120}$ &--\\ 
\hline\hline

\hline
\end{tabular} 
\begin{tablenotes}[para,flushleft]
Notes: all kinematic parameters are given in \kms. Velocity shifts $\Delta V_i$ and line widths FWHM$_i$ refer to the individual Gaussian components ($i = $ 1 to 3, at maximum) required to model the emission line profiles. $V10$ is defined as 10th-percentile velocity of the fitted line profile, while $W80$ is defined as the difference between the 90th- and 10th-percentiles. $v_{max}$ is defined as $\Delta V_{broad} + 2\sigma_{broad}$. 
  \end{tablenotes}
  \end{threeparttable}
\label{tab:kinematics}
\end{table}

\section{Discussion}\label{sec:discussion}


We used the aperture-extracted spectra of the obscured nuclei E and W and the bright cluster NE to identify all gas emission line features in the NIRSpec spectral range, 
as well as the main absorption features due to dust and stars.
We identified $\sim 70$ ionised gas and $\sim 50$ molecular emission lines; Tables \ref{tab:linelist}, \ref{tab:linelist2}, and \ref{tab:linelist3} give the (vacuum) wavelengths and fluxes for all the firmly identified features in Arp220, while Table \ref{tab:kinematics} displays the kinematic parameters of the main optical lines discussed in this paper. We note that a few well detected emission lines are not tabulated, as (i) their modelling would require detailed deblending from other lines, or (ii) their observed wavelengths are very close to the edges of the NIRSpec detector gaps.  

In this section we discuss the main spectral features observed in the spectra reported in Fig. \ref{fig:totalspectra} in relation with the possible presence of intense episodes of starburst activity and AGN accretion in the two nuclei.

\subsection{Star-formation tracers}\label{sec:SFR}

We derived the instantaneous ($\sim 10$ Myr) SFR from the \Paa dust corrected luminosity, following \citet[][and assuming a \citealt{Chabrier2003} initial mass function]{Gimenez2022}: 

\begin{equation}\label{eq:SFR}
\begin{split}
SFR\ [M_\odot \ yr^{-1}] & = 4.4\times 10^{-42} \times \left( 
\frac{H\alpha}{Pa\alpha} \right)_{int} \times L(Pa\alpha)_{corr}\ [erg s^{-1}]\\ & = 3.7 \times 10^{-41} \times L(Pa\alpha)_{corr}\ [erg s^{-1}]
\end{split}
\end{equation}

where the dust-corrected \Paa luminosity can be obtained with A(\Paa) $ = k($\Paa$)\times$ E(B-V). In deriving Eq. \ref{eq:SFR} we assumed electron density of 10$^3$ cm$^{-3}$ (the dependence from n$_e$ is very weak) and temperature of 10$^4$ K. Moreover, for the E and W nuclear emission we consider  only the narrow components (dashed blue Gaussian lines in Fig. \ref{fig:outflows}), thus excluding the contribution of outflowing material.


The use of E(B-V) inferred from \Pab/\Brg imply dust corrections of $\sim 5-8\times$ factors for the \Paa flux, resulting in a SFR $\sim 0.5$ \Msunyr for the W nucleus, $\sim 0.2$ \Msunyr for the E nucleus, and $\sim 0.9$ \Msunyr for the NE cluster. 
Therefore, these three regions account for a total SFR(\Paa) $\sim 2$ \Msunyr. Considering the extraction regions of $\sim 0.01$ kpc$^2$, these values would be translated in star-formation surface densities $\Sigma_{SFR} \sim 20-100$ \Msunyr kpc$^{-2}$.  

We also use the calibration of SFR for the 3.3 $\mu$m PAH from \citet[][see their Eq. 5]{Kim2012} to convert the measured luminosities (L(PAH) $\sim 2\times 10^{40}$ \ergs for the W, L(PAH) $7\times 10^{39}$ \ergs for the E, and L(PAH) $10^{40}$ \ergs for NE) in L(IR), and then the \citet{Kennicutt2012}\footnote{Slightly smaller (but consistent) values would be obtained using the \citealt{PiquerasLopez2016} relation.} relation to derive dust-corrected SFR(PAH) in the range $\sim 1.5-3.4$ \Msunyr (see Table \ref{tab:properties}). 

In previous optical and NIR studies of Arp 220, slightly higher SFR measurements were obtained when integrating over the whole galaxy: \citet{Perna2020} reported SFR(\ha) $\lesssim 10$ \Msunyr (using the Balmer decrement to correct for extinction), while \citet{Gimenez2022} obtained SFR(\Pab) $\sim 19$ \Msunyr from HST narrow-band images (using \Pab/\ha for the dust-correction). \citet{Gimenez2022} also reported the total SFR inferred from the 24 $\mu$m luminosity, 76 \Msunyr; this value is significantly less affected by dust (A$_V \sim 90$ would imply a correction of a factor 2 at 24 $\mu$m). This value is more consistent with the total SFR(PAH) $\sim 40$ \Msunyr we measure from NIRSpec data using an integrated spectrum extracted over a very large aperture ($r = 1.5\arcsec$, i.e. 0.56 kpc).

Therefore, our NIRSpec-based outcomes are definitely in line with previous estimates obtained from optical, NIR, and mid-infrared bands. However, it is worth noting that the SFR inferred from the radio and sub-mm/mm regime appears to be several times higher, reaching 200-250 \Msunyr (e.g. \citealt{Varenius2016, Pereira2021}). In particular, these authors reported SFR $\sim 60-80$ \Msunyr and $\sim 110-150$ \Msunyr for the E and W nuclei separately, hence significantly higher than our NIR-based nuclear measurements. This discrepancy may stem from a combination of factors: (i) The far-IR and radio emission might not be solely attributable to star formation, but could also include contributions from AGN, which are totally extinguished in the NIR. 
(ii) The SFR derived from FIR measurements traces the star formation history over a longer period (up to 100 Myr) compared to the \Paa and PAH features ($\lesssim 10$ Myr). 

Summarising, the NIRSpec-based SFR measurements do not offer a solution to the existing discrepancy among various star-formation tracers identified in previous studies. 

\subsection{High electron densities}\label{sec:density_discussion}

The \feii-based electron densities at the position of the two nuclei and the NE cluster (log (\Ne/cm$^{-3}$) $\sim 3.7-4.5$) consistently exceed those inferred from optical \sii 6718, 32 $\AA$ lines, which are of the order of 200 cm$^{-3}$ (\citealt{Perna2020}). This suggests that \feii lines, having critical densities higher than \sii (n$_{\rm crit}\sim 1600$ cm$^{-3}$ for \sii 6716 $\AA$; n$_{\rm crit}\sim 10^4$ cm$^{-3}$ for iron lines), trace post-shock regions with higher compression.  An alternative explanation might be that the higher spatial resolution of NIRSpec data ($\sim 0.1\arcsec$ at the wavelengths of the \Ne-sensitive \feii lines) compared to the MUSE observations ($\sim 0.6\arcsec$) enables a better isolation of regions with higher degree of ionisation. 

At higher redshift (up to $z\sim 9$), \sii and \oii 3727, 30 $\AA$ line ratios are systematically used to measure the ISM electron densities; values in the range $\sim 300-1000$ cm$^{-3}$ are generally reported for both AGN (e.g. \citealt{Perna2017b,Cresci2023}) and star-forming galaxies (e.g. \citealt{Schreiber2019, Davies2021,RodriguezdelPino2023, Isobe2023}). Significant correlations between  $\Sigma_{SFR}$ and \Ne \ have been reported in the literature (e.g. \citealt{Shimakawa2015}), although they cover relatively narrow ranges in both $\Sigma_{SFR}$ ($\sim$ 0.1 -- 1 \Msunyr kpc$^{-2}$) and \Ne \ ($\sim$ 10 -- 10$^3$ cm$^{-3}$). From the extrapolation of these relations (e.g. Eq. 3 in \citealt{Reddy2023}) to the very high $\Sigma_{SFR}$ measured in Arp 220 (see Sect. \ref{sec:SFR}), we would expect log(n$_e$/cm$^{-3}$) $\sim 3$. However, this value stands roughly 30 times lower than the one measured at the position of the Arp 220 W nucleus using \feii lines.
Instead, our measurements more closely resemble electron densities observed in supernova remnants (e.g., \citealt{Lee2017}) and at the bases of protostar jets (e.g., \citealt{Davis2011}), with \feii-based densities spanning log (\Ne/cm$^{-3}$) $\approx 3.5-4.5$.

Therefore, on the basis of available information, we cannot discriminate between the two above mentioned scenarios: the extremely high electron densities we measured in the nuclei and the NE cluster could be due to 
higher levels of compression and/or ionisation with respect to \sii gas emitting in the optical. We can definitely exclude any potential contamination from high-density AGN Broad Line Region (BLR) in the \feii -based electron densities obtained for the Arp 220 nuclei, as the forbidden \feii line transitions cannot originate from BLR
high-density regions ($N_{\rm e} ({\rm BLR}) \gg  N_c$(\feii)).

\subsection{Presence of multi-phase outflows}\label{sec:outflow_discussion}

The NIRSpec nuclear spectra of Arp220 show ionised and molecular gas transitions with broad profiles and prominent blue wings associated with outflows. In particular, high-velocity components are detected in \Paa and \feii 1.644 $\mu$m at the position of the E and W nuclei. Prominent H$_2$ blue wings are solely observed in the E nucleus. This suggests that the high-velocity gas in the W nucleus is sufficiently hot (because of shocks, or intense ultraviolet radiation by hot stars or AGN) to disassociate the H$_2$ molecules. 

\begin{figure}[!htb]
\centering 
\includegraphics[width=0.46\textwidth]{{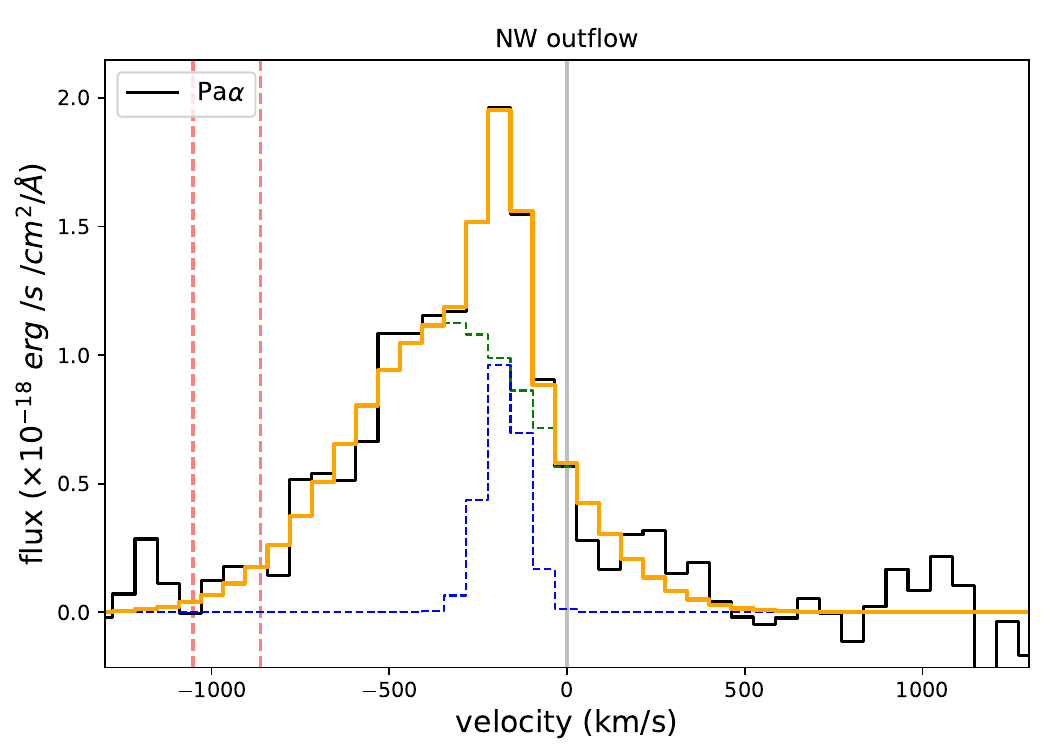}}
\includegraphics[width=0.46\textwidth]{{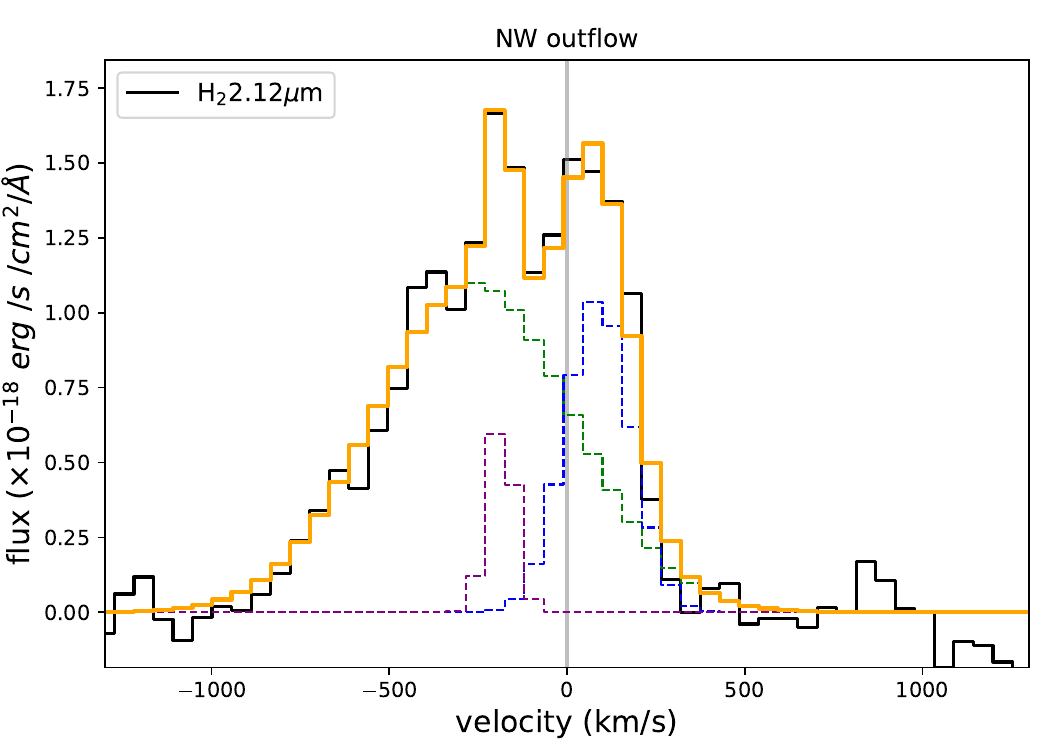}}

\caption{ {\it Observed profiles of \Paa and H$_2$ 2.12 $\mu$m, showing fast outflows.} The spectra are extracted from the region in Fig. \ref{fig:fig1} marked with a cyan diamond, and reported in velocity space. The best-fit emission line profiles are shown in orange; the individual Gaussian profiles used to reproduce the line shapes are reported in blue, purple, and green. The vertical grey lines mark the zero-velocity of the E nucleus (corresponding to the kinematic centre of Arp 220); the dashed red vertical lines in the top panel identify the \hei lines close to \Paa (see Table \ref{tab:linelist}).}\label{fig:fastoutflow}
\end{figure}

The nuclear material can reach velocities of up to $v_{max} \sim 550$ \kms (Fig. \ref{fig:outflows}).  We anticipate here that such outflows reach distances of $\sim 1$ kpc and velocities of $-900$ \kms in the off-nuclear regions covered by NIRSpec (Ulivi et al., in prep.). Figure \ref{fig:fastoutflow} shows the \Paa and H$_2$ 2.12 $\mu$m line profiles extracted from a circular aperture of $r = 0.15\arcsec$ at a distance of $\sim 1.3\arcsec$ to the west from the W nucleus, as defined in Fig. \ref{fig:fig1} (cyan diamond); at this position, both ionised and molecular gas have outflow velocities up to $900$ \kms (Table \ref{tab:kinematics}).
Similar velocities have also been observed in the ionised and atomic gas phases at kpc-scales in the optical regime (e.g. \citealt{Perna2020}). 


Nuclear outflows in NIR \hi, H$_2$ and iron lines have been observed in nearby ULIRGs and active galaxies. While sources lacking an active SMBH typically exhibit less extreme outflow velocities (on the order of a few hundred \kms), AGN can manifest outflows across a broad velocity range, from a few hundred to over a thousand \kms (e.g. \citealt{Emonts2017, Perna2021, Speranza2022, Villar2023}). Consequently, distinguishing between starburst- and AGN-driven outflows in Arp 220, based solely on velocity estimates, proves to be a challenging task. A more comprehensive analysis of the multi-phase and multi-scale outflows will be necessary to elucidate whether the outflow energetics in Arp 220 are more consistent with AGN or starburst launching mechanisms; this work will be presented in a forthcoming paper, Ulivi et al., in prep..







\subsection{Absence of high-ionisation lines}

High excitation emission lines with ionisation potentials IP $> 54$~eV (the He$^+$ ionisation energy) can be formed either in gas photoionised by hard ultraviolet radiation or in very hot, collisionally ionised plasma. Therefore, they can be used as sign for AGN activity (e.g. \citealt{Oliva1994, Moorwood1997}). 

Most of these highly ionised lines have previously been identified in ground-based (e.g. \citealt{RodriguezArdila2011, Lamperti2017, Cerqueira2021, denBrok2022}) and space-based (e.g. \citealt{Lutz2000, Sturm2002}) NIR spectra of some nearby Seyfert galaxies (Table \ref{tab:highion}). 
No quantitative studies of the coronal lines prevalence in the AGN population have been made yet; however, there are indications for a more likely presence of such lines in unobscured type-1 sources rather than in obscured sources. This is consistent with the fact that some coronal lines have been recently detected with NIRSpec in the nearby Seyfert 1.5 nucleus of NGC7469 (\citealt{Bianchin2023}) but not in the highly obscured active nuclei of VV114 (see Appendix \ref{sec:app2}; but see also \citealt{Speranza2022, Alvarez2023}). 
The lack of high ionisation lines in AGN sources may reflect an AGN ionising continuum lacking photons below a few keV: this could explain the absence of such line emission in obscured sources (e.g. \citealt{RodriguezArdila2011}).

High ionisation lines are not detected in the nuclear spectra of Arp 220. However, we note the presence of a bright feature near $\sim 4.488\mu$m, hence close to expected wavelength of the \mgiv line (IP $\sim 80$ eV). Previous studies reported the detection of the \mgiv line in a few nearby active galaxies alongside other highly ionised species; for instance, \cite{Sturm2002} reported \mgiv/[Ar\,{\sc{vi}}] $\sim 0.5$ in Circinus and NGC1068; \citet{Bianchin2023} obtained \mgiv/[Ar\,{\sc{vi}}] $\sim 1.5$. 
Notably, [Ar\,{\sc{vi}}] is not detected in Arp 220. Given its similar IP to \mgiv (see Table \ref{tab:highion}), the observed feature at the expected wavelength of \mgiv may involve alternative transitions. For instance, the low-ionisation line SI ($\lambda 4.4855$ in vacuum, IP $\sim 10$ eV) could be responsible of the line features in the Arp 220 spectra.

\begin{figure}[!htb]
\centering 
\includegraphics[width=0.5\textwidth,trim= 40 0 0 0,clip]{{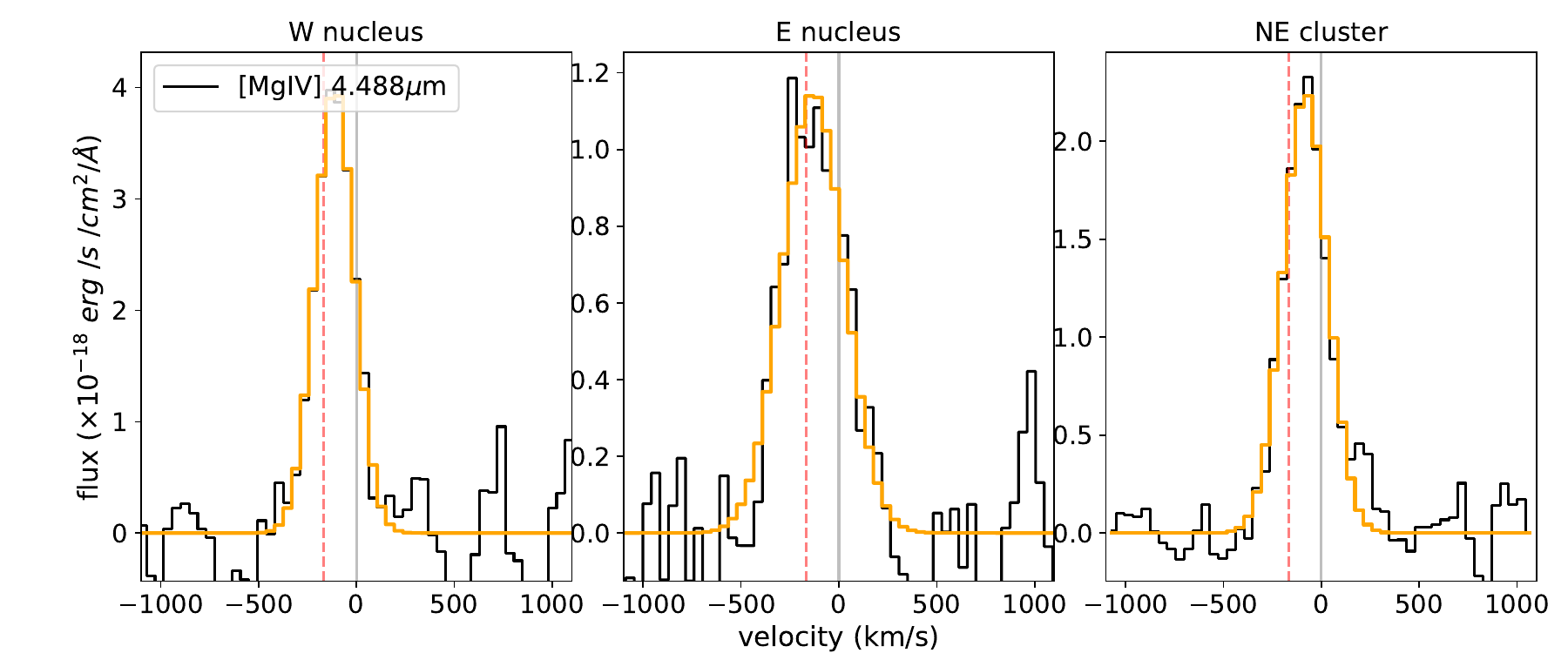}}

\caption{ {\it Continuum-subtracted integrated spectra of the three regions of interest in the vicinity of the \mgiv 4.488 $\mu$m.} 
The original spectra (in black) are reported in velocity space. The best-fit emission line profiles are shown in orange. Broad and asymmetric profiles are found in the nuclear spectra. The vertical grey lines mark the zero-velocity as inferred from the narrow component of \hi lines; the dashed red lines marks the expected position of the S I line.}\label{fig:mgiv}
\end{figure}

Figure \ref{fig:mgiv} shows the emission line at the wavelength position of the \mgiv in the three spectra of interest: these features are broad (FWHM $\sim 300$ \kms) and blue-shifted ($\Delta V\sim -100$ \kms) with respect to the \mgiv expected wavelength (grey solid line at $v=0$ km/s). Moreover, their peaks are not at the systemic of the SI transition neither. Pereira-Santaella et al. (subm.) show that this feature can be associated with shocked \mgiv emission due to supernova explosions. In support of this explanation we note that this line feature is bright at the position of intense clusters in Arp 220, but it is faint in the surroundings of the two nuclei. 

In Table \ref{tab:highion}, we report the main high ionisation lines in the range covered by our NIRSpec spectra; for all but \mgiv we report the $3\sigma$ upper limits inferred from NIRSpec data. These upper limits, and in particular the \sivi one, can be used to compare the NIRSpec measurements with those from the literature. In fact, \sivi is detected in the largest number of AGN spectra (\citealt{Lamperti2017,denBrok2022}). 

\begin{figure}[!tb]
\centering 
\includegraphics[width=0.45\textwidth]{{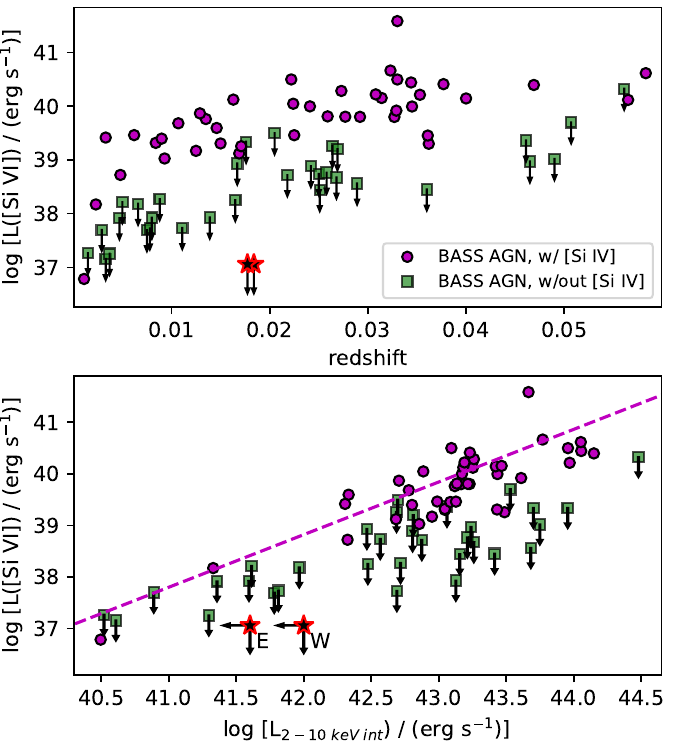}}

\caption{{\it \sivi luminosity for Arp 220 and a sample of nearby AGN.} (Top panel:) \sivi luminosity as a function of redshift for the sample of BASS AGN with detected  (purple circles) and undetected (green squares) \sivi line, from \citet{Lamperti2017}. Arp 220 measurements are represented by red stars. For sources with undetected \sivi lines, we provide three-sigma upper limits (illustrated with arrows). (Bottom panel): \sivi luminosity as a function of intrinsic X-ray (2--10 keV) luminosity for the same sources reported in the top panel. The dashed line indicate the linear relation between L(\sivi) and L$_{2-10\ keV \ int}$ obtained by \citet[][not considering upper limits]{Lamperti2017}.  X-ray luminosity upper limits for the nuclei of Arp 220 are taken from \citet{Paggi2017}.}\label{fig:SiVI}
\end{figure}

Figure \ref{fig:SiVI}, top panel, shows the distribution of L(\sivi) as a function of the redshift for a sample of nearby active galaxies selected in the hard X-ray band (14-195 keV) from the {\it Swift}/Burst Alert Telescope (BAT) survey, as part of the BASS sample (\citealt{Koss2017, Ricci2017}). We display both \sivi  detections and 3$\sigma$ upper limits, alongside the Arp 220 non detections. It is evident that NIRSpec enables us to obtain more stringent upper limits compared to previous observations of other systems at $z \sim 0.018$ (by $\sim 1$ dex). 

Figure \ref{fig:SiVI}, bottom panel, shows the correlation between intrinsic X-ray and \sivi luminosity (e.g. \citealt{RodriguezArdila2011, Lamperti2017}) for the same sources. The upper limits for \sivi in Arp 220 nuclei are indicated at the intrinsic X-ray (2--10 keV) luminosities of $<1\times 10^{42}$ \ergs (W) and $<4\times 10^{41}$ \ergs (E), computed by \citet{Paggi2017} from the $3\sigma$ upper limits on the neutral Fe–K$\alpha$ line. 
The Arp 220 \sivi upper limits are approximately 1 dex below the L(\sivi) -- L$_{\rm 2-10 \ keV}$ relation 
(dashed magenta line). 
However, it should be noted that measuring intrinsic X-ray luminosity for highly obscured sources can be highly uncertain, and that the Arp 220 X-ray luminosities are merely upper limits. Consequently, it is possible that the nuclei of Arp 220 may not exhibit notable distinctions from other AGN with undetected \sivi reported in the figure, for which the presence of accreting SMBHs is confirmed in the hard X-ray band.

To conclude, there is no obvious highly ionised line emission associated with AGN from either nucleus of Arp 220, despite the many bright \hi, \hei, H$_2$, and low excitation \feii lines detected at very high signal-to-noise. Moreover, the very stringent upper limits derived from NIRSpec observations cannot be used to discard the presence of AGN in Arp 220 nuclei when we take into account the extremely high nuclear extinction (Sect. \ref{sec:extinction}). 



\begin{table}[h]
\tabcolsep 3.5pt 
\centering
\caption{High ionisation lines}%
  \begin{threeparttable}
\begin{tabular}{lcclc}
\hline
line & $\lambda_{vac}$ & IP & ref & log $L_{line}$  \\
  &  ($\mu$m) & (eV) & & (\ergs)\\
  {\footnotesize (1)} & {\footnotesize (2)}& {\footnotesize (3)}& {\footnotesize (4)} & {\footnotesize (5)} \\
\hline
$[$S {\sc{viii}}] & 0.9916 & 281 & {\footnotesize L17, C21} & $<$37.0\\
Cr {\sc{viii}} & 1.0109 & 160 & & $<$36.9\\
$[$Fe {\sc{xiii}}] & 1.0750 & 331 &{\footnotesize L17, C21} & $<$37.0\\ 
$[$Fe {\sc{xiii}}] & 1.0801 & 331 &{\footnotesize L17, C21} & $<$37.0\\
$[$S {\sc{ix}}] & 1.2523 & 329 & {\footnotesize B23, C21} & $<$36.9\\
$[$Si {\sc{x}}] & 1.4305 & 351 & {\footnotesize C21} & --\\
Ti {\sc{vi}} & 1.7156 & 99 & & $<$37.3\\
$[$S {\sc{xi}}] & 1.9201 & 448 &{\footnotesize L17, C21} & $<$37.1\\
$[$Si {\sc{xi}}] & 1.9327 & 401 & {\footnotesize L17, C21} & $<$37.1\\
\sivi  & 1.9646 & 167 &{\footnotesize L17, C21, B23} & $<$37.1\\
$[$Al {\sc{ix}}] & 2.0450 & 285 & {\footnotesize C21} & $<$37.1\\
Sc {\sc{v}} & 2.3118 & 73 & & $<$37.0\\
$[$Ca {\sc{viii}}] & 2.3211 & 127 & {\footnotesize C21,R06} & $<$37.0  \\
$[$Si {\sc{vii}}] & 2.4810 & 205 &  {\footnotesize B23} & $<$37.0\\
$[$Si {\sc{ix}}$]$ & 2.5850 & 303 & {\footnotesize S02, L00} & $<$37.1\\
Al {\sc{v}} & 2.9050 & 120 & &  $<$37.0\\
$[$Mg {\sc{viii}}$]$ & 3.0280 & 225 & {\footnotesize S02,B23, L00} & $<$37.0\\
$[$Ca {\sc{iv}}] & 3.2070 & 51 & {\footnotesize L00} & $<$36.7\\
$[$Al {\sc{vi}}] & 3.6603 & 154 & & $<$36.9\\
$[$Al {\sc{viii}}$]$ & 3.6900 & 242 & & $<$36.9 \\
$[$Si {\sc{ix}}] & 3.9290 & 303 & {\footnotesize S02, L00}& $<$37.0 \\
$[$Ca {\sc{v}}] & 4.1585 & 67 & & $<$37.0\\
\mgiv & 4.4880 & 80 & {\footnotesize S02, B23, L00} & $38.03\pm 0.01$ (W) \\
 & & & & $37.72\pm 0.02$ (E) \\
\arvi & 4.5292 & 75 & {\footnotesize B23, L00} & $<$37.0 \\
\hline
\end{tabular} 
\begin{tablenotes}[para,flushleft]
Notes: references of papers reporting the detection of high ionisation lines: \citet[][S02]{Sturm2002}, \citet[][L17]{Lamperti2017}, \citet[][B23]{Bianchin2023}, \citet[][C21]{Cerqueira2021}, \citet[][L00]{Lutz2000}. 
\mgiv luminosities are reported for both E and W nuclei; for all other high ionisation lines we report 3$\sigma$ upper limits for the luminosities, assuming FWHM $= 250$ \kms. \siten is not covered by NIRSpec observations (it falls in detector gap).
  \end{tablenotes}
  \end{threeparttable}
\label{tab:highion}
\end{table}

\subsection{Near-infrared emission line diagnostic diagrams}

We used the diagnostic diagrams presented by \citet{Riffel2013}, \citet{Colina2015}, and \citet{Calabro2023}, to investigate the ionisation mechanisms responsible of the emission line ratios observed in the nuclei and the NE cluster. Prior to analysis, all fluxes required for the diagnostics were corrected for dust extinction, using the $A_V$ values reported in Table \ref{tab:properties}.

According to the diagram presented by \citet{Riffel2013}, based on the H$_2$ 2.12$\mu$m/Br$\gamma$ vs. \feii 1.26$\mu$m/Pa$\beta$ line ratios, both nuclei and the NE cluster are in the AGN  region; the NE cluster has slightly lower line ratios and it is therefore closer to the `star-forming' region of the diagram.
In the diagram proposed by \citet{Colina2015}, based on the H$_2$ 2.12$\mu$m/Br$\gamma$ vs. \feii1.64$\mu$m/Br$\gamma$, the two nuclei and NE cluster are in the region dominated by AGN and supernovae.
Therefore, a similar ionisation mechanism should be present in the W, E, and NE regions of Arp 220 according to both \citet{Riffel2013} and \citet{Colina2015}.

We also explored the diagnostic diagram recently proposed by \citet{Calabro2023}, that use combinations of ratios of the \siii, \feii, \pii and \ci lines over hydrogen recombination lines.
In all the diagrams, the two nuclei are slightly above the maximum starburst line (by 0.1 -- 0.5~dex). 
The NE clump lies on (or just below) the maximum starburst line. We note that the two nuclei are well above the maximum starburst line in the \ci/\Pab diagram, compatible with both AGN and shock ionisations \citep[see Fig. 13 in ][]{Calabro2023}.

In summary, these diagnostics do not allow us to distinguish between supernovae and AGN ionisation mechanisms. 
More detailed  investigations of the flux ratios in the circum-nuclear region of Arp 220 will be presented in a future work.

\section{Conclusions}\label{sec:conclusions}

This paper provides an overview of the JWST/NIRSpec IFS observations of Arp 220. It introduces the data reduction, and the continuum
and emission line data modelling. It shows the near-infrared spectra of the E and W nuclei, in comparison with that of a bright stellar cluster at $\sim 0.4\arcsec$ (0.15 kpc) north-east from the E nucleus. These spectra are extracted using circular apertures of radius 55 pc (0.15\arcsec) for each region. In the following, we summarise our main findings.

We identify broadening and multiple kinematic components in \hi, H$_2$ and \feii lines caused by outflows (Fig. \ref{fig:outflows}). The emission line features show maximum velocities of up to $\sim 550$ \kms in the nuclear spectra. Even higher velocities ($\sim 900$ \kms) are detected in the off-nuclear regions (Fig. \ref{fig:fastoutflow}). However, they do not conclusively represent a direct evidence of AGN activity. A companion paper will report the spatially resolved gas kinematics, and the outflow energetics will be used to test the theoretical predictions for both starburst- and AGN-driven winds (Ulivi et al., in prep.).

High ionisation lines associated with AGN activity are not detected in the nuclear spectra of Arp 220. We observe the presence of bright \mgiv 4.488 $\mu$m lines (80 eV) in the two nuclear spectra and the NE stellar cluster (Fig. \ref{fig:mgiv}). However, no other high ionisation lines are detected; in particular the \arvi 4.529 $\mu$m line, which has an IP and wavelength close to those of \mgiv, is not detected. The \mgiv lines are broader (FWHM $\sim 300$ \kms) and blue-shifted ($\Delta V \sim -100$ \kms) compared to those of the \hi lines; moreover, \mgiv is bright at the position of clusters, but it is faint in the surroundings of the two
nuclei. 
These arguments suggest that this emission line is not produced by an AGN, but rather due to stellar activity. This result is in line with recent findings of Pereira-Santaella et al. (subm.), who recently reported the presence of \mgiv in other nearby ULIRGs not hosting accreting SMBHs, with kinematic properties similar to those in Arp 220; they also provided grids of photoionization and shock models to determine the conditions where \mgiv originates, concluding that shocks by supernova explosions can explain the observed \mgiv lines.

The luminosity upper limits inferred for the high ionisation (IP $= 167$ eV) \sivi emission line at the position of the Arp 220 nuclei are $\sim 1$ dex below the L(\sivi) -- L$_{2-10\ keV, int}$ observed for X-ray AGN (\citealt{RodriguezArdila2011, Lamperti2017}), but still broadly consistent with other sources hosting accreting SMBHs but missing \sivi emission (Fig. \ref{fig:SiVI}). 

We also report the presence of high excitation $^{12}$CO gas transitions (up to J$_{low} = 23$) in the Arp 220 nuclei (Fig. \ref{fig:portion6}), possibly due to AGN activity (e.g. \citealt{GonzalezAlfonso2023}). A companion paper (Buiten et al., in prep.) will provide detailed modelling of CO gas conditions to assess which mechanism among star formation and AGN contributes to the observed high excitation CO lines. 

To conclude, an unambiguous identification of the presence of AGN in the Arp 220 system remains elusive. At the same time, we cannot exclude the presence of accreting SMBHs, because of the extremely high extinction measured at the position of the two nuclei, with recombination lines-based A$_V\sim 11-14$. NIRSpec observations may only probe the outer gaseous layers, and the total extinction towards the core of the Arp 220 nuclear regions could be much higher (see e.g. \citealt{Haas2001, Sakamoto2021a}). This could explain the absence of highly ionised gas transitions, as also reported in previous works on obscured AGN (e.g. \citealt{RodriguezArdila2011, Lamperti2017}).

\section*{Acknowledgements}

We are grateful to Paul van der Werf, Victorine Buiten, and Ang\`ele Taillard for discussing various aspects of this work

MP, SA, and BRP acknowledge support from the research project PID2021-127718NB-I00 of the Spanish Ministry of Science and Innovation/State Agency of Research (MCIN/AEI/10.13039/501100011033). 
IL acknowledges support from PID2022-140483NB-C22 funded by AEI 10.13039/501100011033 and BDC 20221289 funded by MCIN by the Recovery, Transformation and Resilience Plan from the Spanish State, and by NextGenerationEU from the European Union through the Recovery and Resilience Facility. MPS acknowledges support from grant RYC2021-033094-I funded by MICIU/AEI/10.13039/501100011033 and the European Union NextGenerationEU/PRTR. 
RM acknowledges support by the Science and Technology Facilities Council (STFC), by the ERC Advanced Grant 695671 ``QUENCH'', and by the UKRI Frontier Research grant RISEandFALL; RM is further supported by a research professorship from the Royal Society.
GC, MP, and LU acknowledge the support of the INAF Large Grant 2022 "The metal circle: a new sharp view of the baryon cycle up to Cosmic Dawn with the latest generation IFU facilities".

\bibliographystyle{aa} 
\bibliography{aanda.bib}

\begin{thebibliography}{95}
\expandafter\ifx\csname natexlab\endcsname\relax\def\natexlab#1{#1}\fi

\bibitem[{{{\'A}lvarez-M{\'a}rquez} {et~al.}(2023){{\'A}lvarez-M{\'a}rquez},
  {Labiano}, {Guillard}, {Dicken}, {Argyriou}, {Patapis}, {Law}, {Kavanagh},
  {Larson}, {Gasman}, {Mueller}, {Alberts}, {Brandl}, {Colina},
  {Garc{\'\i}a-Mar{\'\i}n}, {Jones}, {Noriega-Crespo}, {Shivaei}, {Temim}, \&
  {Wright}}]{Alvarez2023}
{{\'A}lvarez-M{\'a}rquez}, J., {Labiano}, A., {Guillard}, P., {et~al.} 2023,
  \aap, 672, A108

\bibitem[{{Armus} {et~al.}(2007){Armus}, {Charmandaris}, {Bernard-Salas},
  {Spoon}, {Marshall}, {Higdon}, {Desai}, {Teplitz}, {Hao}, {Devost}, {Brandl},
  {Wu}, {Sloan}, {Soifer}, {Houck}, \& {Herter}}]{Armus2007}
{Armus}, L., {Charmandaris}, V., {Bernard-Salas}, J., {et~al.} 2007, \apj, 656,
  148

\bibitem[{{Arp}(1966)}]{Arp1966}
{Arp}, H. 1966, \apjs, 14, 1

\bibitem[{{Arribas} {et~al.}(2001){Arribas}, {Colina}, \&
  {Clements}}]{Arribas2001}
{Arribas}, S., {Colina}, L., \& {Clements}, D. 2001, \apj, 560, 160

\bibitem[{{Baan} \& {Haschick}(1995)}]{Baan1995}
{Baan}, W.~A. \& {Haschick}, A.~D. 1995, \apj, 454, 745

\bibitem[{{Baba} {et~al.}(2022){Baba}, {Imanishi}, {Izumi}, {Kawamuro},
  {Nguyen}, {Nakagawa}, {Isobe}, {Onishi}, \& {Matsumoto}}]{Baba2022}
{Baba}, S., {Imanishi}, M., {Izumi}, T., {et~al.} 2022, \apj, 928, 184

\bibitem[{{Barcos-Mu{\~n}oz} {et~al.}(2015){Barcos-Mu{\~n}oz}, {Leroy},
  {Evans}, {Privon}, {Armus}, {Condon}, {Mazzarella}, {Meier}, {Momjian},
  {Murphy}, {Ott}, {Reichardt}, {Sakamoto}, {Sanders}, {Schinnerer},
  {Stierwalt}, {Surace}, {Thompson}, \& {Walter}}]{BarcosMunoz2015}
{Barcos-Mu{\~n}oz}, L., {Leroy}, A.~K., {Evans}, A.~S., {et~al.} 2015, \apj,
  799, 10

\bibitem[{{Bennett} {et~al.}(2014){Bennett}, {Larson}, {Weiland}, \&
  {Hinshaw}}]{Bennett2014}
{Bennett}, C.~L., {Larson}, D., {Weiland}, J.~L., \& {Hinshaw}, G. 2014, \apj,
  794, 135

\bibitem[{{Bianchin} {et~al.}(2023){Bianchin}, {U}, {Song}, {Lai}, {Remigio},
  {Barcos-Munoz}, {Diaz-Santos}, {Armus}, {Inami}, {Larson}, {Evans}, {Boker},
  {Kader}, {Linden}, {Charmandaris}, {Malkan}, {Rich}, {Bohn}, {Medling},
  {Stierwalt}, {Mazzarella}, {Law}, {Privon}, {Aalto}, {Appleton}, {Brown},
  {Buiten}, {Finnerty}, {Hayward}, {Howell}, {Iwasawa}, {Kemper}, {Marshall},
  {McKinney}, {Muller-Sanchez}, {Murphy}, {van der Werf}, {Sanders}, \&
  {Surace}}]{Bianchin2023}
{Bianchin}, M., {U}, V., {Song}, Y., {et~al.} 2023, arXiv e-prints,
  arXiv:2308.00209

\bibitem[{B{\"o}ker {et~al.}(2022)B{\"o}ker, Arribas, Lützgendorf,
  de~Oliveira, Beck, Birkmann, Bunker, Charlot, de~Marchi, Ferruit, Giardino,
  Jakobsen, Kumari, L{\'{o}}pez-Caniego, Maiolino, Manjavacas, Marston,
  Moseley, Muzerolle, Ogle, Pirzkal, Rauscher, Rawle, Rix, Sabbi, Sargent,
  Sirianni, te~Plate, Valenti, Willott, \& Zeidler}]{Boker2022}
B{\"o}ker, T., Arribas, S., Lützgendorf, N., {et~al.} 2022, A\&A, 661, A82

\bibitem[{{B{\"o}ker} {et~al.}(2023){B{\"o}ker}, {Beck}, {Birkmann},
  {Giardino}, {Keyes}, {Kumari}, {Muzerolle}, {Rawle}, {Zeidler}, {Abul-Huda},
  {de Oliveira}, {Arribas}, {Bechtold}, {Bhatawdekar}, {Bonaventura}, {Bunker},
  {Cameron}, {Carniani}, {Charlot}, {Curti}, {Espinoza}, {Ferruit}, {Franx},
  {Jakobsen}, {Karakla}, {L{\'o}pez-Caniego}, {L{\"u}tzgendorf}, {Maiolino},
  {Manjavacas}, {Marston}, {Moseley}, {Ogle}, {Perna}, {Pe{\~n}a-Guerrero},
  {Pirzkal}, {Plesha}, {Proffitt}, {Rauscher}, {Rix}, {Rodr{\'\i}guez del
  Pino}, {Rustamkulov}, {Sabbi}, {Sing}, {Sirianni}, {te Plate}, {{\'U}beda},
  {Wahlgren}, {Wislowski}, {Wu}, \& {Willott}}]{Boker2023}
{B{\"o}ker}, T., {Beck}, T.~L., {Birkmann}, S.~M., {et~al.} 2023, \pasp, 135,
  038001

\bibitem[{{Buiten} {et~al.}(2023){Buiten}, {van der Werf}, {Viti}, {Armus},
  {Barr}, {Barcos-Mu{\~n}oz}, {Evans}, {Inami}, {Linden}, {Privon}, {Song},
  {Rich}, {Aalto}, {Appleton}, {B{\"o}ker}, {Charmandaris}, {Diaz-Santos},
  {Hayward}, {Lai}, {Medling}, {Ricci}, \& {U}}]{Buiten2023}
{Buiten}, V.~A., {van der Werf}, P.~P., {Viti}, S., {et~al.} 2023, arXiv
  e-prints, arXiv:2312.01945

\bibitem[{{Calabr{\`o}} {et~al.}(2023){Calabr{\`o}}, {Pentericci}, {Feltre},
  {Arrabal Haro}, {Radovich}, {Seill{\'e}}, {Oliva}, {Daddi}, {Amor{\'\i}n},
  {Bagley}, {Bisigello}, {Buat}, {Castellano}, {Cleri}, {Dickinson},
  {Fern{\'a}ndez}, {Finkelstein}, {Giavalisco}, {Grazian}, {Hathi},
  {Hirschmann}, {Juneau}, {Kartaltepe}, {Koekemoer}, {Lucas}, {Papovich},
  {P{\'e}rez-Gonz{\'a}lez}, {Pirzkal}, {Santini}, {Trump}, {de la Vega},
  {Wilkins}, {Yung}, {Cassata}, {Gobat}, {Mascia}, {Napolitano}, \&
  {Vulcani}}]{Calabro2023}
{Calabr{\`o}}, A., {Pentericci}, L., {Feltre}, A., {et~al.} 2023, \aap, 679,
  A80

\bibitem[{{Cappellari}(2017)}]{Cappellari2017}
{Cappellari}, M. 2017, \mnras, 466, 798

\bibitem[{{Cardelli} {et~al.}(1989){Cardelli}, {Clayton}, \&
  {Mathis}}]{Cardelli1989}
{Cardelli}, J.~A., {Clayton}, G.~C., \& {Mathis}, J.~S. 1989, \apj, 345, 245

\bibitem[{{Cerqueira-Campos} {et~al.}(2021){Cerqueira-Campos},
  {Rodr{\'\i}guez-Ardila}, {Riffel}, {Marinello}, {Prieto}, \&
  {Dahmer-Hahn}}]{Cerqueira2021}
{Cerqueira-Campos}, F.~C., {Rodr{\'\i}guez-Ardila}, A., {Riffel}, R., {et~al.}
  2021, \mnras, 500, 2666

\bibitem[{{Chabrier}(2003)}]{Chabrier2003}
{Chabrier}, G. 2003, \pasp, 115, 763

\bibitem[{{Chandar} {et~al.}(2023){Chandar}, {Caputo}, {Linden}, {Mok},
  {Whitmore}, {Calzetti}, {Elmegreen}, {Lee}, {Ubeda}, {White}, \&
  {Cook}}]{Chandar2023}
{Chandar}, R., {Caputo}, M., {Linden}, S., {et~al.} 2023, \apj, 943, 142

\bibitem[{{Colina} {et~al.}(2004){Colina}, {Arribas}, \&
  {Clements}}]{Colina2004}
{Colina}, L., {Arribas}, S., \& {Clements}, D. 2004, \apj, 602, 181

\bibitem[{{Colina} {et~al.}(2015){Colina}, {Piqueras L{\'o}pez}, {Arribas},
  {Riffel}, {Riffel}, {Rodriguez-Ardila}, {Pastoriza}, {Storchi-Bergmann},
  {Alonso-Herrero}, \& {Sales}}]{Colina2015}
{Colina}, L., {Piqueras L{\'o}pez}, J., {Arribas}, S., {et~al.} 2015, \aap,
  578, A48

\bibitem[{{Cresci} {et~al.}(2023){Cresci}, {Tozzi}, {Perna}, {Brusa},
  {Marconcini}, {Marconi}, {Carniani}, {Brienza}, {Giroletti}, {Belfiore},
  {Ginolfi}, {Mannucci}, {Ulivi}, {Scholtz}, {Venturi}, {Arribas}, {{\"U}bler},
  {D'Eugenio}, {Mingozzi}, {Balmaverde}, {Capetti}, {Parlanti}, \&
  {Zana}}]{Cresci2023}
{Cresci}, G., {Tozzi}, G., {Perna}, M., {et~al.} 2023, arXiv e-prints,
  arXiv:2301.11060

\bibitem[{{Davies} {et~al.}(2021){Davies}, {F{\"o}rster Schreiber}, {Genzel},
  {Shimizu}, {Davies}, {Schruba}, {Tacconi}, {{\"U}bler}, {Wisnioski}, {Wuyts},
  {Fossati}, {Herrera-Camus}, {Lutz}, {Mendel}, {Naab}, {Price}, {Renzini},
  {Wilman}, {Beifiori}, {Belli}, {Burkert}, {Chan}, {Contursi}, {Fabricius},
  {Lee}, {Saglia}, \& {Sternberg}}]{Davies2021}
{Davies}, R.~L., {F{\"o}rster Schreiber}, N.~M., {Genzel}, R., {et~al.} 2021,
  \apj, 909, 78

\bibitem[{{Davis} {et~al.}(2011){Davis}, {Cervantes}, {Nisini}, {Giannini},
  {Takami}, {Whelan}, {Smith}, {Ray}, {Chrysostomou}, \& {Pyo}}]{Davis2011}
{Davis}, C.~J., {Cervantes}, B., {Nisini}, B., {et~al.} 2011, \aap, 528, A3

\bibitem[{{den Brok} {et~al.}(2022){den Brok}, {Koss}, {Trakhtenbrot}, {Stern},
  {Cantalupo}, {Lamperti}, {Ricci}, {Ricci}, {Oh}, {Bauer}, {Riffel},
  {Rodr{\'\i}guez-Ardila}, {B{\"a}r}, {Harrison}, {Ichikawa},
  {Mej{\'\i}a-Restrepo}, {Mushotzky}, {Powell}, {Boissay-Malaquin},
  {Stalevski}, {Treister}, {Urry}, \& {Veilleux}}]{denBrok2022}
{den Brok}, J.~S., {Koss}, M.~J., {Trakhtenbrot}, B., {et~al.} 2022, \apjs,
  261, 7

\bibitem[{{D'Eugenio} {et~al.}(2023){D'Eugenio}, {Perez-Gonzalez}, {Maiolino},
  {Scholtz}, {Perna}, {Circosta}, {Uebler}, {Arribas}, {Boeker}, {Bunker},
  {Carniani}, {Charlot}, {Chevallard}, {Cresci}, {Curtis-Lake}, {Jones},
  {Kumari}, {Lamperti}, {Looser}, {Parlanti}, {Rix}, {Robertson}, {Rodriguez
  Del Pino}, {Tacchella}, {Venturi}, \& {Willott}}]{DEugenio2023}
{D'Eugenio}, F., {Perez-Gonzalez}, P., {Maiolino}, R., {et~al.} 2023, arXiv
  e-prints, arXiv:2308.06317

\bibitem[{{Donnan} {et~al.}(2024){Donnan}, {Garc{\'\i}a-Bernete}, {Rigopoulou},
  {Pereira-Santaella}, {Roche}, \& {Alonso-Herrero}}]{Donnan2024}
{Donnan}, F.~R., {Garc{\'\i}a-Bernete}, I., {Rigopoulou}, D., {et~al.} 2024,
  \mnras [\eprint[arXiv]{2402.17479}]

\bibitem[{{Emonts} {et~al.}(2017){Emonts}, {Colina}, {Piqueras-L{\'o}pez},
  {Garcia-Burillo}, {Pereira-Santaella}, {Arribas}, {Labiano}, \&
  {Alonso-Herrero}}]{Emonts2017}
{Emonts}, B.~H.~C., {Colina}, L., {Piqueras-L{\'o}pez}, J., {et~al.} 2017,
  \aap, 607, A116

\bibitem[{{Engel} {et~al.}(2011){Engel}, {Davies}, {Genzel}, {Tacconi},
  {Sturm}, \& {Downes}}]{Engel2011}
{Engel}, H., {Davies}, R.~I., {Genzel}, R., {et~al.} 2011, \apj, 729, 58

\bibitem[{{Eriksen} {et~al.}(2009){Eriksen}, {Arnett}, {McCarthy}, \&
  {Young}}]{Eriksen2009}
{Eriksen}, K.~A., {Arnett}, D., {McCarthy}, D.~W., \& {Young}, P. 2009, \apj,
  697, 29

\bibitem[{{Fluetsch} {et~al.}(2021){Fluetsch}, {Maiolino}, {Carniani},
  {Arribas}, {Belfiore}, {Bellocchi}, {Cazzoli}, {Cicone}, {Cresci}, {Fabian},
  {Gallagher}, {Ishibashi}, {Mannucci}, {Marconi}, {Perna}, {Sturm}, \&
  {Venturi}}]{Fluetsch2021}
{Fluetsch}, A., {Maiolino}, R., {Carniani}, S., {et~al.} 2021, \mnras, 505,
  5753

\bibitem[{{F{\"o}rster Schreiber} {et~al.}(2019){F{\"o}rster Schreiber},
  {{\"U}bler}, {Davies}, {Genzel}, {Wisnioski}, {Belli}, {Shimizu}, {Lutz},
  {Fossati}, {Herrera-Camus}, {Mendel}, {Tacconi}, {Wilman}, {Beifiori},
  {Brammer}, {Burkert}, {Carollo}, {Davies}, {Eisenhauer}, {Fabricius},
  {Lilly}, {Momcheva}, {Naab}, {Nelson}, {Price}, {Renzini}, {Saglia},
  {Sternberg}, {van Dokkum}, \& {Wuyts}}]{Schreiber2019}
{F{\"o}rster Schreiber}, N.~M., {{\"U}bler}, H., {Davies}, R.~L., {et~al.}
  2019, \apj, 875, 21

\bibitem[{{Garc{\'\i}a-Bernete} {et~al.}(2024){Garc{\'\i}a-Bernete},
  {Pereira-Santaella}, {Gonz{\'a}lez-Alfonso}, {Rigopoulou}, {Efstathiou},
  {Donnan}, \& {Thatte}}]{GarciaBernete2024}
{Garc{\'\i}a-Bernete}, I., {Pereira-Santaella}, M., {Gonz{\'a}lez-Alfonso}, E.,
  {et~al.} 2024, \aap, 682, L5

\bibitem[{{Gibb} {et~al.}(2004){Gibb}, {Whittet}, {Boogert}, \&
  {Tielens}}]{Gibb2004}
{Gibb}, E.~L., {Whittet}, D.~C.~B., {Boogert}, A.~C.~A., \& {Tielens},
  A.~G.~G.~M. 2004, \apjs, 151, 35

\bibitem[{{Gim{\'e}nez-Arteaga} {et~al.}(2022){Gim{\'e}nez-Arteaga}, {Brammer},
  {Marchesini}, {Colina}, {Bajaj}, {Brinch}, {Calzetti}, {Lange-Vagle},
  {Murphy}, {Perna}, {Piqueras-L{\'o}pez}, \& {Snyder}}]{Gimenez2022}
{Gim{\'e}nez-Arteaga}, C., {Brammer}, G.~B., {Marchesini}, D., {et~al.} 2022,
  \apjs, 263, 17

\bibitem[{{Gonz{\'a}lez-Alfonso} {et~al.}(2023){Gonz{\'a}lez-Alfonso},
  {Garc{\'\i}a-Bernete}, {Pereira-Santaella}, {Neufeld}, {Fischer}, \&
  {Donnan}}]{GonzalezAlfonso2023}
{Gonz{\'a}lez-Alfonso}, E., {Garc{\'\i}a-Bernete}, I., {Pereira-Santaella}, M.,
  {et~al.} 2023, arXiv e-prints, arXiv:2312.04914

\bibitem[{{Gustafsson} {et~al.}(2008){Gustafsson}, {Edvardsson}, {Eriksson},
  {J{\o}rgensen}, {Nordlund}, \& {Plez}}]{Gustafsson2008}
{Gustafsson}, B., {Edvardsson}, B., {Eriksson}, K., {et~al.} 2008, \aap, 486,
  951

\bibitem[{{Haas} {et~al.}(2001){Haas}, {Klaas}, {M{\"u}ller}, {Chini}, \&
  {Coulson}}]{Haas2001}
{Haas}, M., {Klaas}, U., {M{\"u}ller}, S.~A.~H., {Chini}, R., \& {Coulson}, I.
  2001, \aap, 367, L9

\bibitem[{{Isobe} {et~al.}(2023){Isobe}, {Ouchi}, {Nakajima}, {Harikane},
  {Ono}, {Xu}, {Zhang}, \& {Umeda}}]{Isobe2023}
{Isobe}, Y., {Ouchi}, M., {Nakajima}, K., {et~al.} 2023, \apj, 956, 139

\bibitem[{Jakobsen {et~al.}(2022)Jakobsen, Ferruit, de~Oliveira, Arribas,
  Bagnasco, Barho, Beck, Birkmann, Böker, Bunker, Charlot, de~Jong, de~Marchi,
  Ehrenwinkler, Falcolini, Fels, Franx, Franz, Funke, Giardino, Gnata, Holota,
  Honnen, Jensen, Jentsch, Johnson, Jollet, Karl, Kling, Köhler, Kolm, Kumari,
  Lander, Lemke, L{\'{o}}pez-Caniego, Lützgendorf, Maiolino, Manjavacas,
  Marston, Maschmann, Maurer, Messerschmidt, Moseley, Mosner, Mott, Muzerolle,
  Pirzkal, Pittet, Plitzke, Posselt, Rapp, Rauscher, Rawle, Rix, Rödel,
  Rumler, Sabbi, Salvignol, Schmid, Sirianni, Smith, Strada, te~Plate, Valenti,
  Wettemann, Wiehe, Wiesmayer, Willott, Wright, Zeidler, \&
  Zincke}]{Jakobsen2022}
Jakobsen, P., Ferruit, P., de~Oliveira, C.~A., {et~al.} 2022, A\&A, 661, A80

\bibitem[{{Kennicutt} \& {Evans}(2012)}]{Kennicutt2012}
{Kennicutt}, R.~C. \& {Evans}, N.~J. 2012, \araa, 50, 531

\bibitem[{{Kim} {et~al.}(2012){Kim}, {Im}, {Lee}, {Lee}, {Jun}, {Nakagawa},
  {Matsuhara}, {Wada}, {Oyabu}, {Takagi}, {Inami}, {Ohyama}, {Yamada}, {Helou},
  {Armus}, \& {Shi}}]{Kim2012}
{Kim}, J.~H., {Im}, M., {Lee}, H.~M., {et~al.} 2012, \apj, 760, 120

\bibitem[{{Koo} {et~al.}(2016){Koo}, {Raymond}, \& {Kim}}]{Koo2016}
{Koo}, B.-C., {Raymond}, J.~C., \& {Kim}, H.-J. 2016, Journal of Korean
  Astronomical Society, 49, 109

\bibitem[{{Koss} {et~al.}(2017){Koss}, {Trakhtenbrot}, {Ricci}, {Lamperti},
  {Oh}, {Berney}, {Schawinski}, {Balokovi{\'c}}, {Baronchelli}, {Crenshaw},
  {Fischer}, {Gehrels}, {Harrison}, {Hashimoto}, {Hogg}, {Ichikawa}, {Masetti},
  {Mushotzky}, {Sartori}, {Stern}, {Treister}, {Ueda}, {Veilleux}, \&
  {Winter}}]{Koss2017}
{Koss}, M., {Trakhtenbrot}, B., {Ricci}, C., {et~al.} 2017, \apj, 850, 74

\bibitem[{{Lamperti} {et~al.}(2017){Lamperti}, {Koss}, {Trakhtenbrot},
  {Schawinski}, {Ricci}, {Oh}, {Landt}, {Riffel}, {Rodr{\'\i}guez-Ardila},
  {Gehrels}, {Harrison}, {Masetti}, {Mushotzky}, {Treister}, {Ueda}, \&
  {Veilleux}}]{Lamperti2017}
{Lamperti}, I., {Koss}, M., {Trakhtenbrot}, B., {et~al.} 2017, \mnras, 467, 540

\bibitem[{{Lamperti} {et~al.}(2022){Lamperti}, {Pereira-Santaella}, {Perna},
  {Colina}, {Arribas}, {Garc{\'\i}a-Burillo}, {Gonz{\'a}lez-Alfonso}, {Aalto},
  {Alonso-Herrero}, {Combes}, {Labiano}, {Piqueras-L{\'o}pez}, {Rigopoulou}, \&
  {van der Werf}}]{Lamperti2022}
{Lamperti}, I., {Pereira-Santaella}, M., {Perna}, M., {et~al.} 2022, \aap, 668,
  A45

\bibitem[{{Law} {et~al.}(2023){Law}, {Morrison}, {Argyriou}, {Patapis},
  {{\'A}lvarez-M{\'a}rquez}, {Labiano}, \& {Vandenbussche}}]{Law2023}
{Law}, D.~D., {Morrison}, J.~E., {Argyriou}, I., {et~al.} 2023, \aj, 166, 45

\bibitem[{{Lee} {et~al.}(2017){Lee}, {Koo}, {Moon}, {Burton}, \&
  {Lee}}]{Lee2017}
{Lee}, Y.-H., {Koo}, B.-C., {Moon}, D.-S., {Burton}, M.~G., \& {Lee}, J.-J.
  2017, \apj, 837, 118

\bibitem[{{Lockhart} {et~al.}(2015){Lockhart}, {Kewley}, {Lu}, {Allen},
  {Rupke}, {Calzetti}, {Davies}, {Dopita}, {Engel}, {Heckman}, {Leitherer}, \&
  {Sanders}}]{Lockhart2015}
{Lockhart}, K.~E., {Kewley}, L.~J., {Lu}, J.~R., {et~al.} 2015, \apj, 810, 149

\bibitem[{{Lutz} {et~al.}(2000){Lutz}, {Sturm}, {Genzel}, {Moorwood},
  {Alexander}, {Netzer}, \& {Sternberg}}]{Lutz2000}
{Lutz}, D., {Sturm}, E., {Genzel}, R., {et~al.} 2000, \apj, 536, 697

\bibitem[{{May} \& {Steiner}(2017)}]{May2017}
{May}, D. \& {Steiner}, J.~E. 2017, \mnras, 469, 994

\bibitem[{{McDowell} {et~al.}(2003){McDowell}, {Clements}, {Lamb}, {Shaked},
  {Hearn}, {Colina}, {Mundell}, {Borne}, {Baker}, \& {Arribas}}]{McDowell2003}
{McDowell}, J.~C., {Clements}, D.~L., {Lamb}, S.~A., {et~al.} 2003, \apj, 591,
  154

\bibitem[{{Moorwood} {et~al.}(1997){Moorwood}, {Marconi}, {van der Werf}, \&
  {Oliva}}]{Moorwood1997}
{Moorwood}, A.~F.~M., {Marconi}, A., {van der Werf}, P.~P., \& {Oliva}, E.
  1997, \apss, 248, 113

\bibitem[{{Nardini} {et~al.}(2010){Nardini}, {Risaliti}, {Watabe}, {Salvati},
  \& {Sani}}]{Nardini2010}
{Nardini}, E., {Risaliti}, G., {Watabe}, Y., {Salvati}, M., \& {Sani}, E. 2010,
  \mnras, 405, 2505

\bibitem[{{Oliva} {et~al.}(1995){Oliva}, {Origlia}, {Kotilainen}, \&
  {Moorwood}}]{Oliva1995}
{Oliva}, E., {Origlia}, L., {Kotilainen}, J.~K., \& {Moorwood}, A.~F.~M. 1995,
  \aap, 301, 55

\bibitem[{{Oliva} {et~al.}(1994){Oliva}, {Salvati}, {Moorwood}, \&
  {Marconi}}]{Oliva1994}
{Oliva}, E., {Salvati}, M., {Moorwood}, A.~F.~M., \& {Marconi}, A. 1994, \aap,
  288, 457

\bibitem[{Osterbrock \& Ferland(2006)}]{Osterbrock2006}
Osterbrock, D.~E. \& Ferland, G.~J. 2006, Astrophysics of gaseous nebulae and
  active galactic nuclei (University Science Books)

\bibitem[{{Paggi} {et~al.}(2017){Paggi}, {Fabbiano}, {Risaliti}, {Wang},
  {Karovska}, {Elvis}, {Maksym}, {McDowell}, \& {Gallagher}}]{Paggi2017}
{Paggi}, A., {Fabbiano}, G., {Risaliti}, G., {et~al.} 2017, \apj, 841, 44

\bibitem[{{Pereira-Santaella} {et~al.}(2021){Pereira-Santaella}, {Colina},
  {Garc{\'\i}a-Burillo}, {Lamperti}, {Gonz{\'a}lez-Alfonso}, {Perna},
  {Arribas}, {Alonso-Herrero}, {Aalto}, {Combes}, {Labiano},
  {Piqueras-L{\'o}pez}, {Rigopoulou}, \& {van der Werf}}]{Pereira2021}
{Pereira-Santaella}, M., {Colina}, L., {Garc{\'\i}a-Burillo}, S., {et~al.}
  2021, \aap, 651, A42

\bibitem[{{Pereira-Santaella} {et~al.}(2024){Pereira-Santaella},
  {Gonz{\'a}lez-Alfonso}, {Garc{\'\i}a-Bernete}, {Garc{\'\i}a-Burillo}, \&
  {Rigopoulou}}]{Pereira2024a}
{Pereira-Santaella}, M., {Gonz{\'a}lez-Alfonso}, E., {Garc{\'\i}a-Bernete}, I.,
  {Garc{\'\i}a-Burillo}, S., \& {Rigopoulou}, D. 2024, \aap, 681, A117

\bibitem[{{Perna} {et~al.}(2020){Perna}, {Arribas}, {Catal{\'a}n-Torrecilla},
  {Colina}, {Bellocchi}, {Fluetsch}, {Maiolino}, {Cazzoli}, {Hern{\'a}n
  Caballero}, {Pereira Santaella}, {Piqueras L{\'o}pez}, \& {Rodr{\'\i}guez del
  Pino}}]{Perna2020}
{Perna}, M., {Arribas}, S., {Catal{\'a}n-Torrecilla}, C., {et~al.} 2020, \aap,
  643, A139

\bibitem[{{Perna} {et~al.}(2023){Perna}, {Arribas}, {Marshall}, {D'Eugenio},
  {{\"U}bler}, {Bunker}, {Charlot}, {Carniani}, {Jakobsen}, {Maiolino},
  {Rodr{\'\i}guez Del Pino}, {Willott}, {B{\"o}ker}, {Circosta}, {Cresci},
  {Curti}, {Husemann}, {Kumari}, {Lamperti}, {P{\'e}rez-Gonz{\'a}lez}, \&
  {Scholtz}}]{Perna2023a}
{Perna}, M., {Arribas}, S., {Marshall}, M., {et~al.} 2023, \aap, 679, A89

\bibitem[{{Perna} {et~al.}(2021){Perna}, {Arribas}, {Pereira Santaella},
  {Colina}, {Bellocchi}, {Catal{\'a}n-Torrecilla}, {Cazzoli}, {Crespo
  G{\'o}mez}, {Maiolino}, {Piqueras L{\'o}pez}, \& {Rodr{\'\i}guez del
  Pino}}]{Perna2021}
{Perna}, M., {Arribas}, S., {Pereira Santaella}, M., {et~al.} 2021, \aap, 646,
  A101

\bibitem[{{Perna} {et~al.}(2015){Perna}, {Brusa}, {Cresci}, {Comastri},
  {Lanzuisi}, {Lusso}, {Marconi}, {Salvato}, {Zamorani}, {Bongiorno},
  {Mainieri}, {Maiolino}, \& {Mignoli}}]{Perna2015a}
{Perna}, M., {Brusa}, M., {Cresci}, G., {et~al.} 2015, \aap, 574, A82

\bibitem[{{Perna} {et~al.}(2017){Perna}, {Lanzuisi}, {Brusa}, {Cresci}, \&
  {Mignoli}}]{Perna2017b}
{Perna}, M., {Lanzuisi}, G., {Brusa}, M., {Cresci}, G., \& {Mignoli}, M. 2017,
  \aap, 606, A96

\bibitem[{{Piqueras L{\'o}pez} {et~al.}(2016){Piqueras L{\'o}pez}, {Colina},
  {Arribas}, {Pereira-Santaella}, \& {Alonso-Herrero}}]{PiquerasLopez2016}
{Piqueras L{\'o}pez}, J., {Colina}, L., {Arribas}, S., {Pereira-Santaella}, M.,
  \& {Alonso-Herrero}, A. 2016, \aap, 590, A67

\bibitem[{{Ralchenko}(2005)}]{Ralchenko2005}
{Ralchenko}, Y. 2005, Memorie della Societa Astronomica Italiana Supplementi,
  8, 96

\bibitem[{{Reddy} {et~al.}(2023){Reddy}, {Sanders}, {Shapley}, {Topping},
  {Kriek}, {Coil}, {Mobasher}, {Siana}, \& {Rezaee}}]{Reddy2023}
{Reddy}, N.~A., {Sanders}, R.~L., {Shapley}, A.~E., {et~al.} 2023, \apj, 951,
  56

\bibitem[{{Ricci} {et~al.}(2017){Ricci}, {Trakhtenbrot}, {Koss}, {Ueda}, {Del
  Vecchio}, {Treister}, {Schawinski}, {Paltani}, {Oh}, {Lamperti}, {Berney},
  {Gandhi}, {Ichikawa}, {Bauer}, {Ho}, {Asmus}, {Beckmann}, {Soldi},
  {Balokovi{\'c}}, {Gehrels}, \& {Markwardt}}]{Ricci2017}
{Ricci}, C., {Trakhtenbrot}, B., {Koss}, M.~J., {et~al.} 2017, \apjs, 233, 17

\bibitem[{{Rich} {et~al.}(2023){Rich}, {Aalto}, {Evans}, {Charmandaris},
  {Privon}, {Lai}, {Inami}, {Linden}, {Armus}, {Diaz-Santos}, {Appleton},
  {Barcos-Mu{\~n}oz}, {B{\"o}ker}, {Larson}, {Law}, {Malkan}, {Medling},
  {Song}, {U}, {van der Werf}, {Bohn}, {Brown}, {Finnerty}, {Hayward},
  {Howell}, {Iwasawa}, {Kemper}, {Marshall}, {Mazzarella}, {McKinney},
  {Muller-Sanchez}, {Murphy}, {Sanders}, {Soifer}, {Stierwalt}, \&
  {Surace}}]{Rich2023}
{Rich}, J., {Aalto}, S., {Evans}, A.~S., {et~al.} 2023, \apjl, 944, L50

\bibitem[{{Riffel} {et~al.}(2013){Riffel}, {Rodr{\'\i}guez-Ardila}, {Aleman},
  {Brotherton}, {Pastoriza}, {Bonatto}, \& {Dors}}]{Riffel2013}
{Riffel}, R., {Rodr{\'\i}guez-Ardila}, A., {Aleman}, I., {et~al.} 2013, \mnras,
  430, 2002

\bibitem[{{Riffel} {et~al.}(2019){Riffel}, {Rodr{\'\i}guez-Ardila},
  {Brotherton}, {Peletier}, {Vazdekis}, {Riffel}, {Martins}, {Bonatto}, {Zanon
  Dametto}, {Dahmer-Hahn}, {Runnoe}, {Pastoriza}, {Chies-Santos}, \&
  {Trevisan}}]{Riffel2019}
{Riffel}, R., {Rodr{\'\i}guez-Ardila}, A., {Brotherton}, M.~S., {et~al.} 2019,
  \mnras, 486, 3228

\bibitem[{{Riffel} {et~al.}(2014){Riffel}, {Vale}, {Storchi-Bergmann}, \&
  {McGregor}}]{Riffel2014}
{Riffel}, R.~A., {Vale}, T.~B., {Storchi-Bergmann}, T., \& {McGregor}, P.~J.
  2014, \mnras, 442, 656

\bibitem[{{Rodr{\'\i}guez-Ardila} {et~al.}(2011){Rodr{\'\i}guez-Ardila},
  {Prieto}, {Portilla}, \& {Tejeiro}}]{RodriguezArdila2011}
{Rodr{\'\i}guez-Ardila}, A., {Prieto}, M.~A., {Portilla}, J.~G., \& {Tejeiro},
  J.~M. 2011, \apj, 743, 100

\bibitem[{{Rodr{\'\i}guez Del Pino} {et~al.}(2023){Rodr{\'\i}guez Del Pino},
  {Perna}, {Arribas}, {D'Eugenio}, {Lamperti}, {P{\'e}rez-Gonz{\'a}lez},
  {{\"U}bler}, {Bunker}, {Carniani}, {Charlot}, {Maiolino}, {Willott},
  {B{\"o}ker}, {Chevallard}, {Cresci}, {Curti}, {Jones}, {Parlanti}, {Scholtz},
  \& {Venturi}}]{RodriguezdelPino2023}
{Rodr{\'\i}guez Del Pino}, B., {Perna}, M., {Arribas}, S., {et~al.} 2023, arXiv
  e-prints, arXiv:2309.14431

\bibitem[{{Rothman}(2021)}]{Rothman2021}
{Rothman}, L.~S. 2021, Nature Reviews Physics, 3, 302

\bibitem[{{Sakamoto} {et~al.}(2021{\natexlab{a}}){Sakamoto},
  {Gonz{\'a}lez-Alfonso}, {Mart{\'\i}n}, {Wilner}, {Aalto}, {Evans}, \&
  {Harada}}]{Sakamoto2021a}
{Sakamoto}, K., {Gonz{\'a}lez-Alfonso}, E., {Mart{\'\i}n}, S., {et~al.}
  2021{\natexlab{a}}, \apj, 923, 206

\bibitem[{{Sakamoto} {et~al.}(2021{\natexlab{b}}){Sakamoto}, {Mart{\'\i}n},
  {Wilner}, {Aalto}, {Evans}, \& {Harada}}]{Sakamoto2021b}
{Sakamoto}, K., {Mart{\'\i}n}, S., {Wilner}, D.~J., {et~al.}
  2021{\natexlab{b}}, \apj, 923, 240

\bibitem[{{Schwarz}(1978)}]{Schwarz1978}
{Schwarz}, U.~J. 1978, \aap, 65, 345

\bibitem[{Scoville {et~al.}(2007)Scoville, Aussel, Brusa, Capak, Carollo,
  Elvis, Giavalisco, Guzzo, Hasinger, Impey, Kneib, LeFevre, Lilly, Mobasher,
  Renzini, Rich, Sanders, Schinnerer, Schminovich, Shopbell, Taniguchi, \&
  Tyson}]{Scoville2007}
Scoville, N., Aussel, H., Brusa, M., {et~al.} 2007, ApJS, 172, 1

\bibitem[{{Scoville} {et~al.}(2017){Scoville}, {Murchikova}, {Walter},
  {Vlahakis}, {Koda}, {Vanden Bout}, {Barnes}, {Hernquist}, {Sheth}, {Yun},
  {Sanders}, {Armus}, {Cox}, {Thompson}, {Robertson}, {Zschaechner}, {Tacconi},
  {Torrey}, {Hayward}, {Genzel}, {Hopkins}, {van der Werf}, \&
  {Decarli}}]{Scoville2017}
{Scoville}, N., {Murchikova}, L., {Walter}, F., {et~al.} 2017, \apj, 836, 66

\bibitem[{{Scoville} {et~al.}(1998){Scoville}, {Evans}, {Dinshaw}, {Thompson},
  {Rieke}, {Schneider}, {Low}, {Hines}, {Stobie}, {Becklin}, \&
  {Epps}}]{Scoville1998}
{Scoville}, N.~Z., {Evans}, A.~S., {Dinshaw}, N., {et~al.} 1998, \apjl, 492,
  L107

\bibitem[{{Shimakawa} {et~al.}(2015){Shimakawa}, {Kodama}, {Steidel}, {Tadaki},
  {Tanaka}, {Strom}, {Hayashi}, {Koyama}, {Suzuki}, \&
  {Yamamoto}}]{Shimakawa2015}
{Shimakawa}, R., {Kodama}, T., {Steidel}, C.~C., {et~al.} 2015, \mnras, 451,
  1284

\bibitem[{{Speranza} {et~al.}(2022){Speranza}, {Ramos Almeida},
  {Acosta-Pulido}, {Riffel}, {Tadhunter}, {Pierce}, {Rodr{\'\i}guez-Ardila},
  {Coloma Puga}, {Brusa}, {Musiimenta}, {Alexander}, {Lapi}, {Shankar}, \&
  {Villforth}}]{Speranza2022}
{Speranza}, G., {Ramos Almeida}, C., {Acosta-Pulido}, J.~A., {et~al.} 2022,
  \aap, 665, A55

\bibitem[{{Sturm} {et~al.}(2002){Sturm}, {Lutz}, {Verma}, {Netzer},
  {Sternberg}, {Moorwood}, {Oliva}, \& {Genzel}}]{Sturm2002}
{Sturm}, E., {Lutz}, D., {Verma}, A., {et~al.} 2002, \aap, 393, 821

\bibitem[{{Teng} {et~al.}(2015){Teng}, {Rigby}, {Stern}, {Ptak}, {Alexander},
  {Bauer}, {Boggs}, {Brandt}, {Christensen}, {Comastri}, {Craig}, {Farrah},
  {Gandhi}, {Hailey}, {Harrison}, {Hickox}, {Koss}, {Luo}, {Treister}, \&
  {Zhang}}]{Teng2015}
{Teng}, S.~H., {Rigby}, J.~R., {Stern}, D., {et~al.} 2015, \apj, 814, 56

\bibitem[{{Ueda} {et~al.}(2022){Ueda}, {Michiyama}, {Iono}, {Miyamoto}, \&
  {Saito}}]{Ueda2022}
{Ueda}, J., {Michiyama}, T., {Iono}, D., {Miyamoto}, Y., \& {Saito}, T. 2022,
  \pasj, 74, 407

\bibitem[{{Ulivi}~et al.(in~prep.)}]{Uliviinprep}
{Ulivi}~et al., . in~prep., --, --,

\bibitem[{{Varenius} {et~al.}(2019){Varenius}, {Conway}, {Batejat},
  {Mart{\'\i}-Vidal}, {P{\'e}rez-Torres}, {Aalto}, {Alberdi}, {Lonsdale}, \&
  {Diamond}}]{Varenius2019}
{Varenius}, E., {Conway}, J.~E., {Batejat}, F., {et~al.} 2019, \aap, 623, A173

\bibitem[{{Varenius} {et~al.}(2016){Varenius}, {Conway}, {Mart{\'\i}-Vidal},
  {Aalto}, {Barcos-Mu{\~n}oz}, {K{\"o}nig}, {P{\'e}rez-Torres}, {Deller},
  {Mold{\'o}n}, {Gallagher}, {Yoast-Hull}, {Horellou}, {Morabito}, {Alberdi},
  {Jackson}, {Beswick}, {Carozzi}, {Wucknitz}, \&
  {Ram{\'\i}rez-Olivencia}}]{Varenius2016}
{Varenius}, E., {Conway}, J.~E., {Mart{\'\i}-Vidal}, I., {et~al.} 2016, \aap,
  593, A86

\bibitem[{{Vazdekis} {et~al.}(2016){Vazdekis}, {Koleva}, {Ricciardelli},
  {R{\"o}ck}, \& {Falc{\'o}n-Barroso}}]{Vazdekis2016}
{Vazdekis}, A., {Koleva}, M., {Ricciardelli}, E., {R{\"o}ck}, B., \&
  {Falc{\'o}n-Barroso}, J. 2016, \mnras, 463, 3409

\bibitem[{{Verro} {et~al.}(2022){Verro}, {Trager}, {Peletier}, {Lan{\c{c}}on},
  {Gonneau}, {Vazdekis}, {Prugniel}, {Chen}, {Coelho},
  {S{\'a}nchez-Bl{\'a}zquez}, {Martins}, {Arentsen}, {Lyubenova},
  {Falc{\'o}n-Barroso}, \& {Dries}}]{Verro2022}
{Verro}, K., {Trager}, S.~C., {Peletier}, R.~F., {et~al.} 2022, \aap, 660, A34

\bibitem[{{Villar Mart{\'\i}n} {et~al.}(2023){Villar Mart{\'\i}n},
  {Castro-Rodr{\'\i}guez}, {Pereira Santaella}, {Lamperti}, {Tadhunter},
  {Emonts}, {Colina}, {Alonso Herrero}, {Cabrera-Lavers}, \&
  {Bellocchi}}]{Villar2023}
{Villar Mart{\'\i}n}, M., {Castro-Rodr{\'\i}guez}, N., {Pereira Santaella}, M.,
  {et~al.} 2023, \aap, 673, A25

\bibitem[{{Wheeler} {et~al.}(2020){Wheeler}, {Glenn}, {Rangwala}, \&
  {Fyhrie}}]{Wheeler2020}
{Wheeler}, J., {Glenn}, J., {Rangwala}, N., \& {Fyhrie}, A. 2020, \apj, 896, 43

\bibitem[{{Yoast-Hull} {et~al.}(2017){Yoast-Hull}, {Gallagher}, {Aalto}, \&
  {Varenius}}]{YoastHull2017}
{Yoast-Hull}, T.~M., {Gallagher}, John~S., I., {Aalto}, S., \& {Varenius}, E.
  2017, \mnras, 469, L89

\bibitem[{{Yoast-Hull} \& {Murray}(2019)}]{Yoast2019}
{Yoast-Hull}, T.~M. \& {Murray}, N. 2019, \mnras, 484, 3665

\end{thebibliography}

\appendix

\section{NIRSpec narrow-band images}

\begin{figure}[t]
\centering 

\includegraphics[width=0.45\textwidth]{{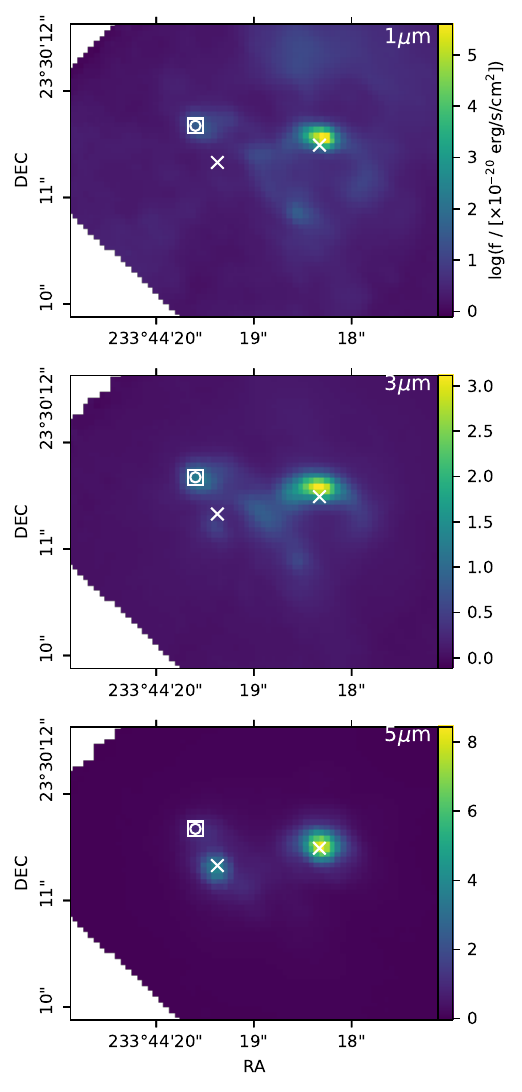}}
{\phantomsubcaption\label{fig:figA1.a}
 \phantomsubcaption\label{fig:figA1.b}
 \phantomsubcaption\label{fig:figA1.c}
}

\caption{ {\it NIRSpec narrow-band images tracing the continuum emission at $\sim 1\mu$m (top), $3\mu$m (centre) and $5\mu$m (bottom).} The position of the W and E nuclei are marked with {\sc x} symbols; the box-circle symbol marks the position of a bright cluster (see also Fig. \ref{fig:fig1}) }\label{fig:figA1}
\end{figure}

\section{Integrated spectra}\label{sec:app2}

\begin{figure*}[!htb]
\centering 
\includegraphics[width=\textwidth]{{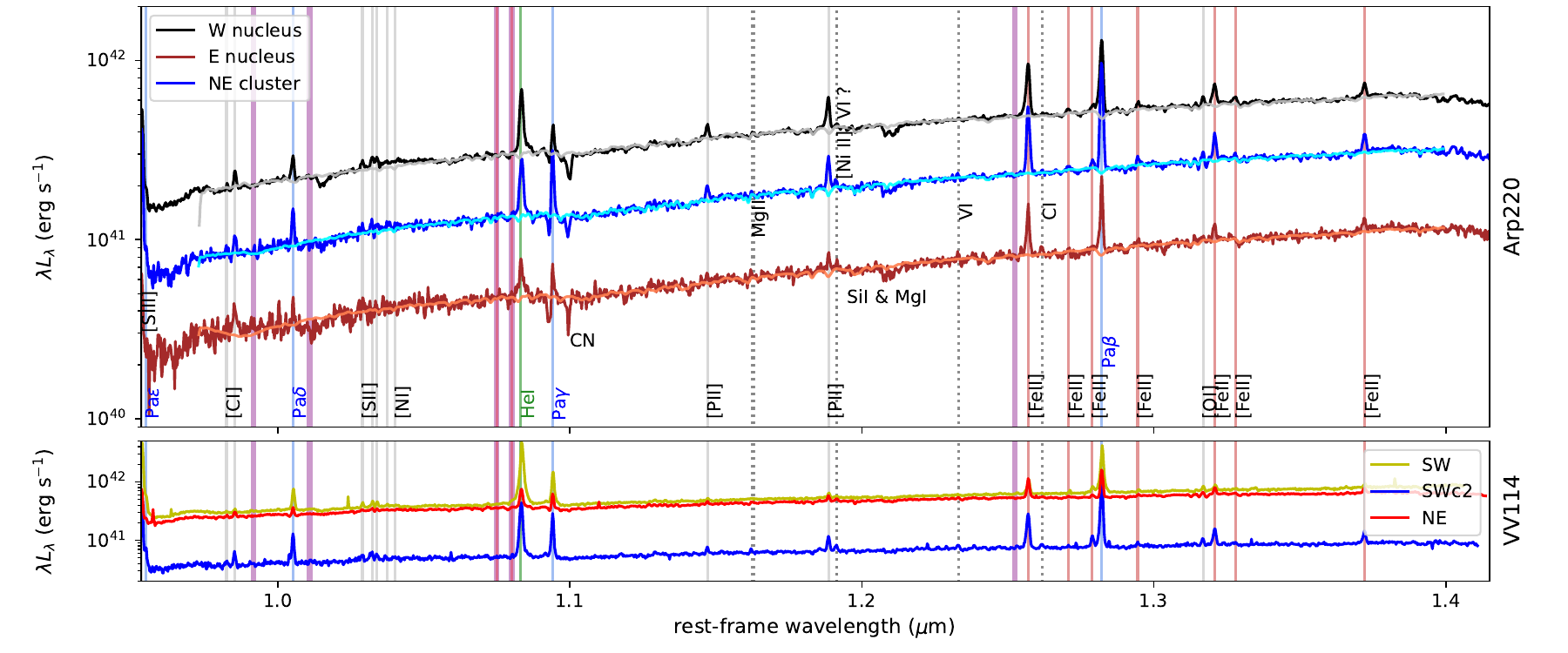}}

\caption{ {\it  $0.95-1.42\ \mu$m portion on the spectra of the Arp 220 and VV114 nuclear regions.} The spectra are shown in luminosity as a function of the rest frame wavelength, considering the redshift of each region. The grey, cyan and orange curves represent the pPXF best-fit. The vertical blue lines mark the position of hydrogen transitions; the orange lines identify H$_2$ lines; the red lines are associated with \feii transitions; the green lines mark \hei features; solid vertical lines mark the position of faint metal lines; potential identifications are indicated with dotted vertical lines. The positions of highly ionised gas transitions are indicated with narrow vertical bands in purple. In the figure, we also indicate the position of strong stellar absorption features. }\label{fig:portion1}
\end{figure*}

\begin{figure*}[!htb]
\centering 
\includegraphics[width=\textwidth]{{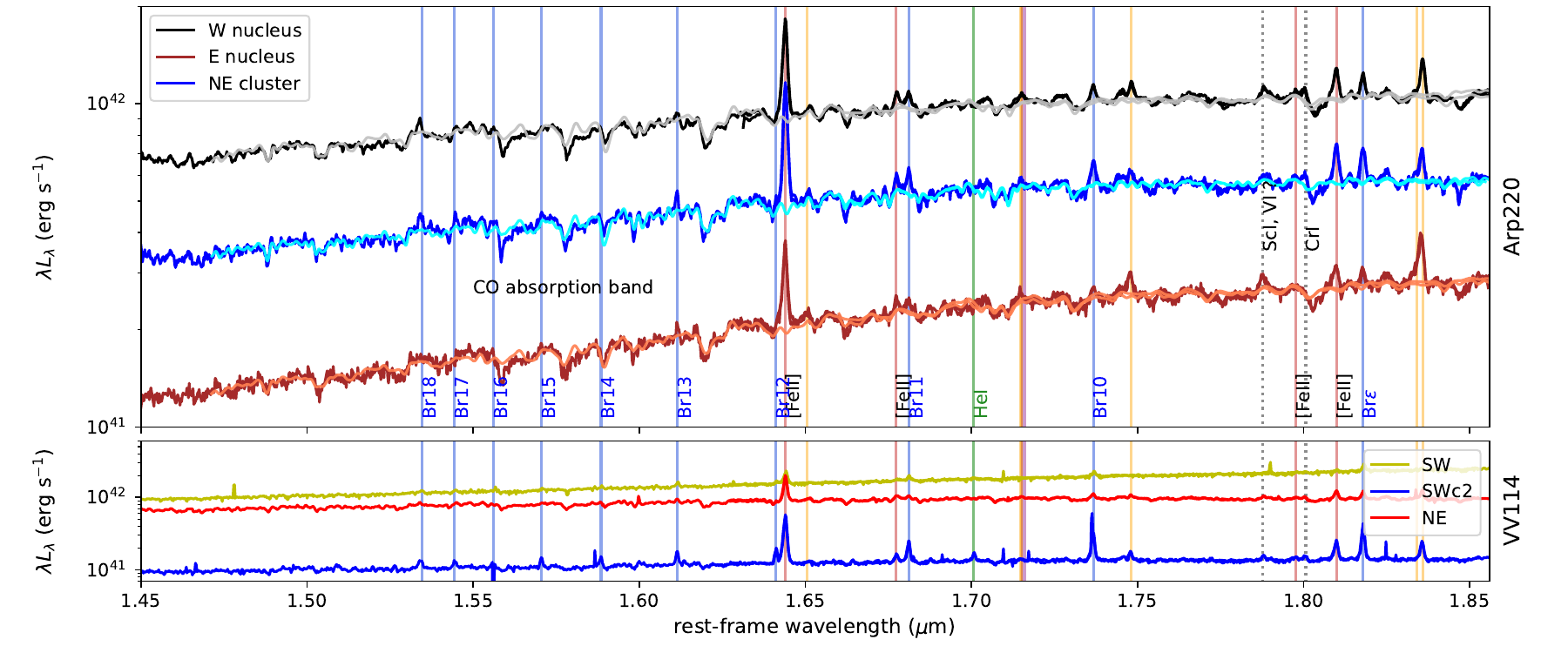}}

\caption{ {\it  $1.45-1.86\ \mu$m portion on the spectra of the Arp 220 and VV114 nuclear regions.} See \ref{fig:portion1} for details. }\label{fig:portion2}
\end{figure*}

\begin{figure*}[!htb]
\centering 
\includegraphics[width=\textwidth]{{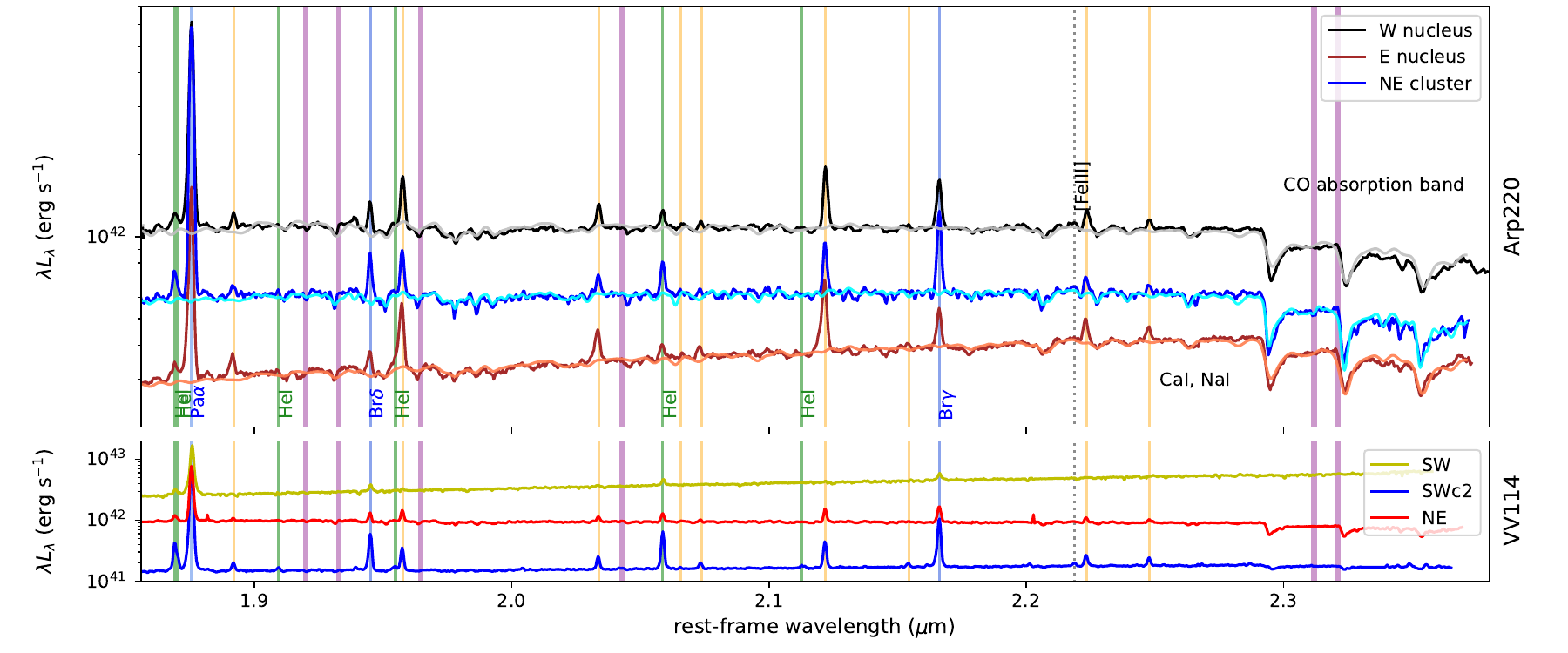}}

\caption{ {\it  $1.85-2.40\ \mu$m portion on the spectra of the Arp 220 and VV114 nuclear regions.} See \ref{fig:portion1} for further details. }\label{fig:portion3}
\end{figure*}

\begin{figure*}[!htb]
\centering 
\includegraphics[width=\textwidth]{{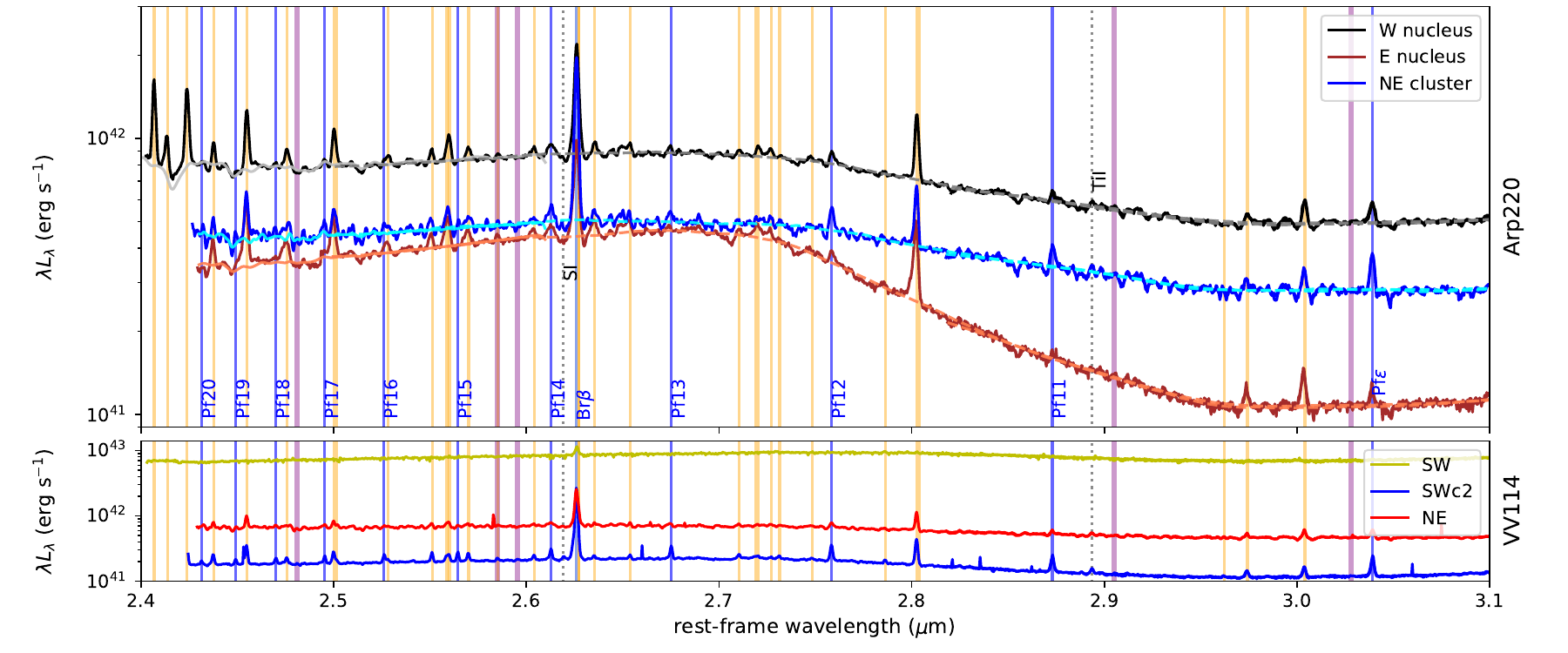}}

\caption{ {\it  $2.4-3.1\ \mu$m portion on the spectra of the Arp 220 and VV114 nuclear regions.} The pseudo-continuum described in Sect. \ref{sec:continuum} are shown with dashed curves. See \ref{fig:portion1} for further details.}\label{fig:portion4}
\end{figure*}

\begin{figure*}[!htb]
\centering 
\includegraphics[width=\textwidth]{{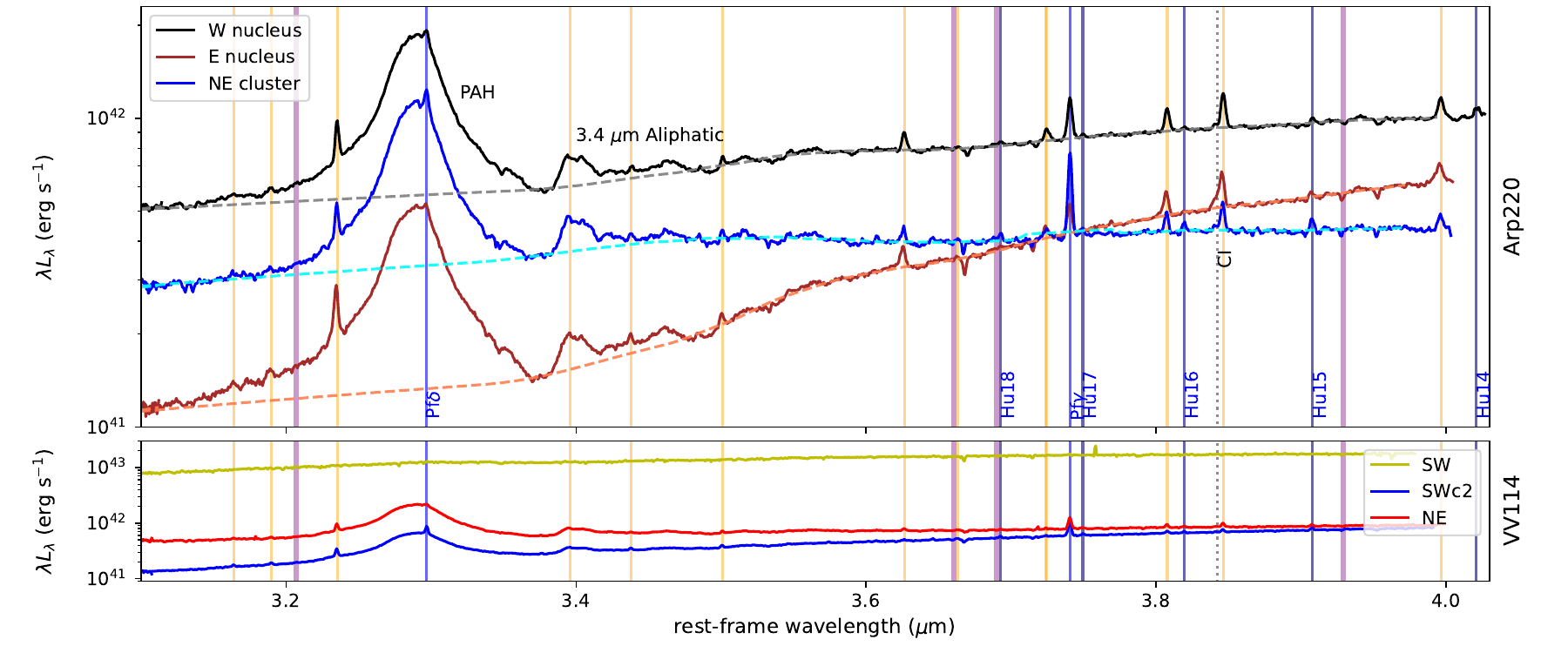}}

\caption{ {\it  $3.1-4.0\ \mu$m portion on the spectra of the Arp 220 and VV114 nuclear regions.} See \ref{fig:portion1} for further details.  }\label{fig:portion5}
\end{figure*}

\begin{figure*}[!htb]
\centering 
\includegraphics[width=\textwidth]{{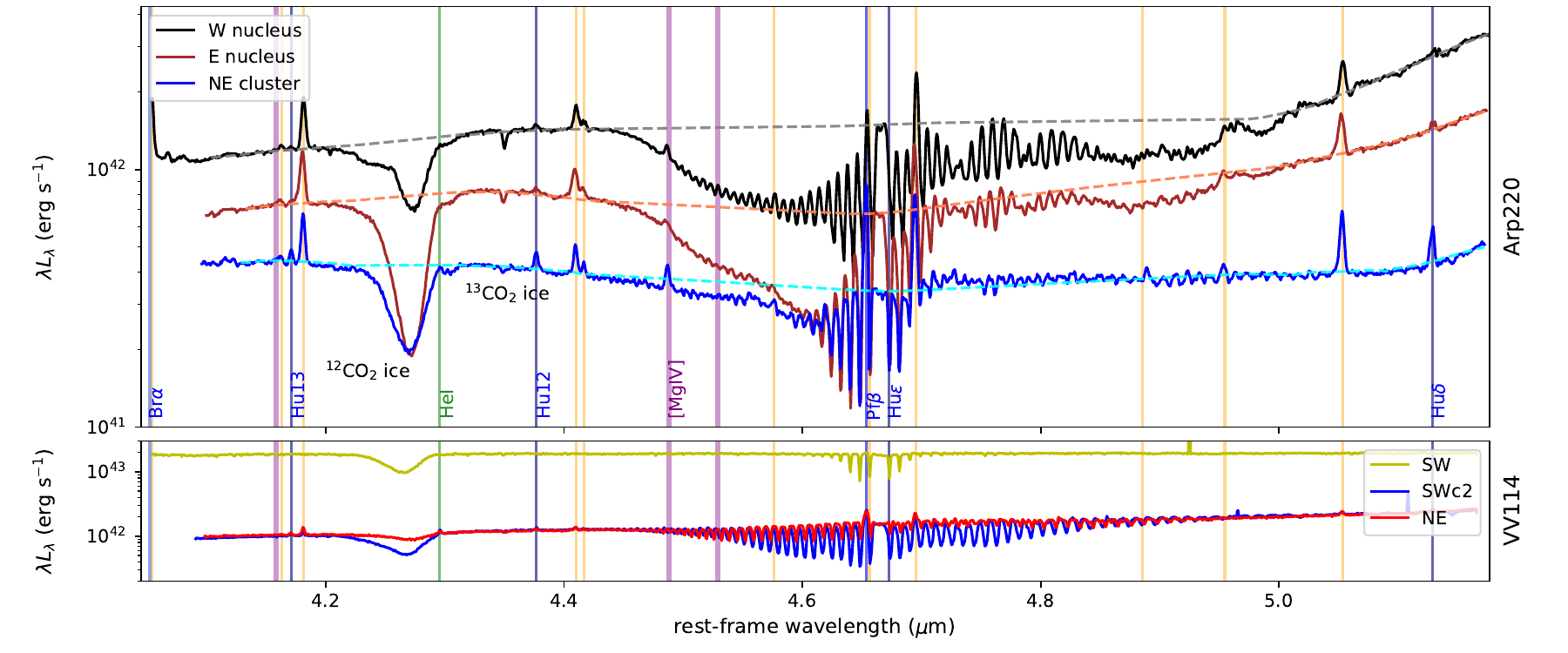}}

\caption{ {\it  $4.1-5.2\ \mu$m portion on the spectra of Arp 220 and VV114 nuclear regions.} The E and W nuclei show $^{12}$CO ro-vibrational lines up to J$_{low} = 23$ (close to the \mgiv line transition, at $\sim 4.48\ \mu$m, and at $\sim 4.88\ \mu$m; see e.g. \citealt{GonzalezAlfonso2023}). Detailed modelling of the CO transitions will be presented in Buiten et al., in prep.. See \ref{fig:portion1} for further details.}\label{fig:portion6}
\end{figure*}

\section{Arp 220 emission line list (band2 and band3 cubes)}

\begin{table*}[h]
\centering
\caption{Arp 220 emission line list in the W and E nuclei, and NE cluster (band2 spectra)}%
\begin{threeparttable}
\begin{tabular}{|lc|c|c|c|}
\hline

\hline
line & $\lambda_{vac}$ & $f_W$ &  $f_E$ & $f_{NE}$ \\
     &  ($\mu$m) &  ($10^{-18}$ \ergs cm$^{-2}$)  &  ($10^{-18}$ \ergs cm$^{-2}$) &  ($10^{-18}$ \ergs cm$^{-2}$)  \\
\hline

\hei & 1.869044 & 273.0$_{-3.3}^{+6.8}$ & 62.3$_{-2.8}^{+4.7}$ & 193.0$_{-3.9}^{+5.0}$ \\
\hei & 1.870234 & 112.0$_{-3.5}^{+5.8}$ & 35.1$_{-2.9}^{+4.6}$ & 36.7$_{-3.3}^{+6.3}$ \\

\Paa & 1.875627 & 7260.0$_{-2.9}^{+5.4}$ & 1510.0$_{-2.6}^{+2.1}$ & 6390.0$_{-4.0}^{+5.8}$ \\
H$_2$ & 1.892094 & 253.0$_{-5.0}^{+5.1}$ & 77.7$_{-11.9}^{+32.3}$ & 85.6$_{-1.0}^{+0.8}$ \\
\Brd & 1.945095 & 455.0$_{-5.7}^{+4.5}$ & 77.2$_{-2.8}^{+4.5}$ & 334.0$_{-3.2}^{+6.6}$ \\

\hei & 1.954840 & 27.2$_{-1.8}^{+3.1}$ & 46.6$_{-3.1}^{+2.7}$ & 30.5$_{-2.7}^{+2.1}$ \\

H$_2$ & 1.957722 & 659.0$_{-5.4}^{+6.9}$ & 243.0$_{-45.6}^{+86.2}$ & 278.0$_{-3.3}^{+8.8}$ \\

H$_2$ & 2.033927 & 268.0$_{-3.4}^{+3.4}$ & 149.0$_{-3.4}^{+5.1}$ & 116.0$_{-3.1}^{+3.8}$ \\

\hei & 2.058600 & 247.0$_{-1.3}^{+1.4}$ & 62.6$_{-0.6}^{+1.3}$ & 215.0$_{-1.1}^{+2.2}$ \\

H$_2$ & 2.065756 & 81.7$_{-0.7}^{+0.5}$ & 28.9$_{-2.0}^{+4.7}$ & 33.3$_{-4.2}^{+4.2}$ \\
H$_2$ & 2.073655 & 126.0$_{-4.6}^{+3.0}$ & 69.6$_{-3.4}^{+4.0}$ & 59.1$_{-4.3}^{+3.1}$ \\

H$_2$ & 2.122011 & 901.0$_{-5.1}^{+7.9}$ & 427.0$_{-3.6}^{+6.0}$ & 345.0$_{-5.4}^{+6.6}$ \\

\Brg & 2.166129 & 762.0$_{-4.1}^{+5.5}$ & 156.0$_{-4.9}^{+9.2}$ & 686.0$_{-2.8}^{+5.9}$ \\

H$_2$ & 2.223475 & 263.0$_{-5.5}^{+3.4}$ & 123.0$_{-4.1}^{+4.2}$ & 100.0$_{-2.8}^{+5.5}$ \\
H$_2$ & 2.247903 & 147.0$_{-3.3}^{+2.4}$ & 76.2$_{-3.5}^{+3.0}$ & 69.6$_{-0.9}^{+1.0}$ \\

H$_2$ & 2.406793 & 1010.0$_{-5.6}^{+9.7}$ & -- & -- \\
H$_2$ & 2.413640 & 402.0$_{-3.9}^{+8.0}$ & -- & -- \\
H$_2$ & 2.423932 & 940.0$_{-4.7}^{+9.2}$ & -- & -- \\

H$_2$ & 2.454956 & 633$_{-3.0}^{+5.4}$ & 375$_{-4.4}^{+5.0}$ & 223$_{-3.8}^{+5.3}$ \\

H$_2$ & 2.475765 & 173.0$_{-1.8}^{+3.1}$ & 124.0$_{-4.0}^{+3.0}$ & 62.4$_{-0.8}^{+1.0}$ \\
H$_2$ & 2.500174 & 332.0$_{-2.1}^{+4.0}$ & 207.0$_{-3.1}^{+4.4}$ & 106.0$_{-2.6}^{+4.9}$ \\
H$_2$ & 2.501650 & 50.9$_{-2.4}^{+3.0}$ & 19.4$_{-2.7}^{+5.0}$ & 39.0$_{-3.1}^{+2.9}$ \\
H$_2$ & 2.528247 & 87.8$_{-1.2}^{+0.6}$ & 58.6$_{-1.7}^{+3.9}$ & 34.6$_{-1.7}^{+3.9}$ \\
H$_2$ & 2.551198 & 103.0$_{-1.6}^{+5.0}$ & 72.7$_{-1.7}^{+4.1}$ & 33.8$_{-2.9}^{+4.1}$ \\
H$_2$ & 2.558725 & 96.0$_{-2.7}^{+4.6}$ & 61.4$_{-3.7}^{+5.3}$ & 54.8$_{-1.7}^{+2.2}$ \\
H$_2$ & 2.560064 & 193.0$_{-2.6}^{+3.3}$ & 121.0$_{-2.6}^{+5.4}$ & 48.9$_{-2.1}^{+3.0}$ \\

H$_2$  & 2.570043 & 124.0$_{-2.7}^{+4.5}$ & 89.6$_{-2.7}^{+3.3}$ & 44.8$_{-2.5}^{+2.8}$ \\
H$_2$  & 2.585185 & 31.3$_{-1.6}^{+2.5}$ & 20.1$_{-3.6}^{+4.1}$ & 1.76$_{-0.7}^{+3.6}$ \\
H$_2$  & 2.604197 & 66.5$_{-2.4}^{+3.7}$ & 46.1$_{-3.0}^{+4.8}$ & 25.5$_{-3.1}^{+4.1}$ \\

Pf 14  & 2.612655 & 121.0$_{-0.1}^{+0.3}$ & 70.9$_{-2.3}^{+4.7}$ & 84.6$_{-2.2}^{+6.0}$ \\
\Brb & 2.625878 & 1840.0$_{-3.9}^{+6.2}$ & 749.0$_{-4.4}^{+6.5}$ & 1610.0$_{-4.8}^{+5.9}$ \\

H$_2$  & 2.635329 & 103.0$_{-3.0}^{+2.9}$ & 89.8$_{-3.2}^{+4.0}$ & 45.8$_{-3.8}^{+3.5}$ \\
H$_2$  & 2.654084 & 80.5$_{-4.3}^{+3.1}$ & 65.1$_{-2.6}^{+3.7}$ & 24.2$_{-1.6}^{+5.8}$ \\

Pf 13 & 2.675139 & 58.9$_{-6.0}^{+5.1}$ & 17.1$_{-3.5}^{+4.0}$ & 51.4$_{-2.5}^{+6.4}$ \\

H$_2$  & 2.710478 & 48.9$_{-3.0}^{+0.4}$ & 19.7$_{-1.4}^{+5.4}$ & 12.7$_{-2.7}^{+3.6}$ \\
H$_2$  & 2.718847 & 1.19$_{-0.004}^{+0.006}$ & 2.29$_{-0.74}^{+5.5}$ & 4.55$_{-2.2}^{+1.6}$ \\
H$_2$  & 2.720371 & 103.0$_{-0.94}^{+2.4}$ & 81.3$_{-2.2}^{+4.9}$ & 26.4$_{-1.9}^{+3.7}$ \\
H$_2$  & 2.726984 & 87.1$_{-2.6}^{+1.0}$ & 59.6$_{-1.5}^{+3.4}$ & 33.2$_{-1.5}^{+3.0}$ \\
H$_2$   & 2.731438 & 17.3$_{-2.6}^{+5.4}$ & 10.0$_{-2.1}^{+2.9}$ & 14.8$_{-1.6}^{+3.6}$ \\
H$_2$   & 2.748299 & 32.0$_{-3.4}^{+4.0}$ & 20.3$_{-2.8}^{+3.4}$ & 11.2$_{-3.9}^{+2.5}$ \\
Pf 12 & 2.758276 & 110.0$_{-1.9}^{+6.3}$ & 48.0$_{-2.7}^{+4.6}$ & 109.0$_{-3.0}^{+4.3}$ \\

H$_2$   & 2.786396 & 12.5$_{-2.0}^{+4.4}$ & 13.6$_{-2.8}^{+4.0}$ & 6.99$_{-2.4}^{+3.7}$ \\
H$_2$  & 2.802750 & 601.0$_{-3.8}^{+3.6}$ & 313.0$_{-1.3}^{+4.3}$ & 244.0$_{-1.7}^{+3.2}$ \\
H$_2$  & 2.804052 & 1.25$_{-0.004}^{+0.007}$ & 1.41$_{-0.02}^{+0.03}$ & 1.08$_{-0.01}^{+0.01}$ \\
Pf 11 & 2.873004 & 58.7$_{-2.4}^{+5.7}$ & 8.35$_{-1.8}^{+3.2}$ & 74.4$_{-2.6}^{+3.5}$ \\

H$_2$  & 2.974311 & 54.0$_{-2.5}^{+2.5}$ & 22.7$_{-2.4}^{+2.7}$ & 27.2$_{-2.2}^{+2.9}$ \\
H$_2$  & 3.004118 & 132.0$_{-2.8}^{+3.7}$ & 48.5$_{-2.1}^{+2.7}$ & 53.5$_{-1.9}^{+3.5}$ \\
\Pfe & 3.039211 & 128.0$_{-2.9}^{+7.3}$ & 21.3$_{-4.1}^{+3.5}$ & 110.0$_{-3.3}^{+6.6}$ \\

\hline\hline

\hline
\end{tabular} 
\begin{tablenotes}[para,flushleft]
Notes: This list comprises all emission lines detected in band2; see Tables \ref{tab:linelist} and \ref{tab:linelist3} for those in band1 and band3, respectively. The reported fluxes are obtained from single Gaussian fits, without dust- and aperture-corrections.     
  \end{tablenotes}
  \end{threeparttable}

\label{tab:linelist2}
\end{table*}

\begin{table*}[h]
\centering
\caption{Arp 220 emission line list in the W and E nuclei, and NE cluster (band3 spectra)}%
\begin{threeparttable}
\begin{tabular}{|lc|c|c|c|}

\hline
line & $\lambda_{vac}$ & $f_W$ &  $f_E$ & $f_{NE}$ \\
     &  ($\mu$m) &  ($10^{-18}$ \ergs cm$^{-2}$)  &  ($10^{-18}$ \ergs cm$^{-2}$) &  ($10^{-18}$ \ergs cm$^{-2}$)  \\
\hline

H$_2$  & 3.164034 & 15.2$_{-2.4}^{+2.1}$ & 8.13$_{-1.6}^{+2.0}$ & 13.4$_{-2.3}^{+3.0}$ \\
H$_2$  & 3.190077 & 28.2$_{-1.8}^{+2.4}$ & 13.4$_{-1.4}^{+1.1}$ & 11.9$_{-2.2}^{+2.6}$ \\
H$_2$  & 3.235257 & 310.0$_{-2.2}^{+3.3}$ & 117.0$_{-1.1}^{+2.8}$ & 117.0$_{-2.2}^{+2.2}$ \\

\Pfd   & 3.297001 & 409.0$_{-0.6}^{+0.8}$ & 74.3$_{-3.5}^{+5.5}$ & 293.0$_{-2.3}^{+1.7}$ \\

H$_2$  & 3.626469 & 152.0$_{-2.7}^{+3.5}$ & 70.4$_{-1.6}^{+2.5}$ & 47.6$_{-1.4}^{+4.1}$ \\
H$_2$  & 3.663464 & 23.2$_{-2.6}^{+2.8}$ & 29.0$_{-1.7}^{+1.9}$ & 12.2$_{-1.0}^{+3.2}$ \\

Hu 18  & 3.692642 & 30.2$_{-1.6}^{+2.3}$ & 9.63$_{-0.8}^{+2.1}$ & 23.5$_{-1.9}^{+2.8}$ \\
H$_2$ & 3.723999 & 5.49$_{-4.7}^{+4.8}$ & 0.898$_{-0.006}^{+0.021}$ & 3.75$_{-2.2}^{+4.8}$ \\
H$_2$  & 3.724736 & 110.0$_{-4.0}^{+4.7}$ & 59.3$_{-1.2}^{+2.6}$ & 30.4$_{-4.3}^{+3.8}$ \\

\Pfg   & 3.740568 & 414.0$_{-3.3}^{+3.8}$ & 111.0$_{-3.5}^{+4.7}$ & 398.0$_{-1.7}^{+2.6}$ \\

Hu 17  & 3.749402 & 12.6$_{-2.1}^{+3.2}$ & 2.87$_{-2.0}^{+2.9}$ & 19.5$_{-2.4}^{+3.1}$ \\
H$_2$ & 3.807736 & 215.0$_{-2.7}^{+2.1}$ & 93.5$_{-16.9}^{+33.1}$ & 70.6$_{-1.6}^{+2.2}$ \\

Hu 16  & 3.819460 & 16.9$_{-2.0}^{+3.3}$ & 3.96$_{-2.1}^{+3.2}$ & 25.4$_{-2.3}^{+2.7}$ \\

C I (?) & 3.841473 & 24.2$_{-0.9}^{+1.7}$ & 26$_{-1.6}^{+2.4}$ & 13.3$_{-1.3}^{+1.9}$ \\

H$_2$  & 3.846434 & 349.0$_{-2.6}^{+3.1}$ & 186.0$_{-2.4}^{+2.9}$ & 99.6$_{-1.2}^{+3.4}$ \\

Hu 15  & 3.907557 & 41.7$_{-2.5}^{+4.3}$ & 23.6$_{-1.7}^{+3.2}$ & 36.7$_{-1.8}^{+2.7}$ \\
H$_2$   & 4.162772 & 63.1$_{-2.5}^{+5.4}$ & 47.0$_{-2.2}^{+3.5}$ & 24.1$_{-1.5}^{+1.7}$ \\
Hu 13  & 4.170803 & 26.8$_{-3.6}^{+3.6}$ & 77.4$_{-10.3}^{+18.8}$ & 45.0$_{-1.9}^{+3.4}$ \\

H$_2$  & 4.181426 & 941.0$_{-2.6}^{+5.4}$ & 580.0$_{-3.3}^{+4.8}$ & 248.0$_{-1.7}^{+3.0}$ \\

Hu 12  & 4.376464 & 78.1$_{-3.0}^{+3.7}$ & 33.0$_{-1.6}^{+3.4}$ & 59.5$_{-1.2}^{+3.4}$ \\
H$_2$   & 4.410159 & 454.0$_{-2.0}^{+2.1}$ & 292.0$_{-1.3}^{+4.9}$ & 117.0$_{-1.2}^{+2.2}$ \\
H$_2$   & 4.416979 & 142.0$_{-1.8}^{+2.8}$ & 71.2$_{-2.3}^{+3.7}$ & 41.4$_{-1.0}^{+1.4}$ \\
H$_2$  & 5.053536 & 972.0$_{-4.8}^{+7.1}$ & 644.0$_{-5.8}^{+5.7}$ & 289.0$_{-1.9}^{+4.5}$ \\
\Hud   & 5.128669 & 148.0$_{-3.6}^{+6.7}$ & 120.0$_{-3.5}^{+5.3}$ & 183.0$_{-2.7}^{+2.1}$ \\

\hline\hline

\hline
\end{tabular} 
\begin{tablenotes}[para,flushleft]
Notes: This list comprises all emission lines detected in band3; see also Tables \ref{tab:linelist} and \ref{tab:linelist2} for those in band1 and band2, respectively. The reported fluxes are obtained from single Gaussian fits, without dust- and aperture-corrections.     
  \end{tablenotes}
  \end{threeparttable}

\label{tab:linelist3}
\end{table*}

\end{document}